\documentclass[prl, aps, amssymb, twocolumn, superscriptaddress, notitlepage, longbibliography]{revtex4-2}

\usepackage{bm,amssymb,amsmath,mathrsfs,amstext,latexsym,physics}
\usepackage{cancel,hhline}
\usepackage[usenames,dvipsnames]{xcolor}
\usepackage[colorlinks=true,citecolor=blue,urlcolor=blue,linkcolor=blue]{hyperref}

\usepackage[caption=false,subrefformat=parens,labelformat=empty]{subfig}
\usepackage{graphicx,tikz}
\usepackage{orcidlink}

\begin{document}
\title{Universal Non-stabilizerness Dynamics Across Quantum Phase Transitions}

\author{Andr\'as Grabarits\orcidlink{0000-0002-0633-7195}}
\email{andras.grabarits@uni.lu}
\affiliation{Department  of  Physics  and  Materials  Science,  University  of  Luxembourg,  L-1511  Luxembourg, Luxembourg}

\author{Adolfo~del Campo\orcidlink{0000-0003-2219-2851}}                 
\affiliation{Department  of  Physics  and  Materials  Science,  University  of  Luxembourg,  L-1511  Luxembourg,  Luxembourg}
\affiliation{Donostia International Physics Center,  E-20018 San Sebasti\'an, Spain}

\date{\today}


\begin{abstract}
 Quantum magic, or non-stabilizerness, is an important quantum resource that characterizes computational power beyond classically simulable Clifford operations and is therefore essential for achieving quantum advantage. While previous studies have explored non-stabilizerness dynamics in random circuits and under time-independent generators, here we extend the study of its universal dynamics to time-dependent driving across quantum phase transitions. In particular, we show that the stabilizer R\'enyi entropies and the cumulants of the Pauli spectrum exhibit universal power-law scaling with the driving rate in slow processes. Moreover, we show that the logarithmic Pauli spectrum is asymptotically Gaussian, implying a lognormal distribution for the Pauli spectrum values. Our results are explicitly demonstrated by exact results in the transverse-field Ising model and by analytical approximations in long-range Kitaev models.
\end{abstract}

\maketitle

\emph{Introduction.---}
Harnessing the intrinsic complexity of quantum systems holds the promise of surpassing classical computational capabilities. Entanglement is often regarded as the primary resource underlying quantum advantage~\cite{Preskill2012}, yet it alone is not sufficient~\cite{JozsaLinden2003_EntanglementNotEnough,Vidal2003_EfficientSimEntangledPRL}. Remarkably, even highly entangled states produced by Clifford operations, known as stabilizer states, can still be efficiently simulated on a classical computer, as established by the Gottesman–Knill theorem~\cite{Chitambar_RevMod2019,Leone_QMisQM2021,Gottesman1998, Gottesman1998_Faulttolerant,Aaronson2004}. This observation naturally calls for different notions of quantum complexity beyond classical simulability. One such resource is non-stabilizerness, or quantum magic, quantifying the non-Clifford operations by which a state deviates from the stabilizer manifold ~\cite{Kitaev_2005, Hamma_PRL_2022,Hamma_Renyi_PRA2024,zhang2024MagicDynamicsRandomCircuit,Turkeshi2025_QuditMagicDynamics,Haug_PRXQ_ScalableMAgic_2023,Howard2014_ContextualityMagicNature}. 
Within the resource-theoretic framework, several monotones have been proposed to quantify non-stabilizerness, including the robustness of magic, discrete Wigner negativity, mana, and further geometric constructions~\cite{HammaNonInt2025,Liu_MBMagic_2022,Haug_PRXQ_ScalableMAgic_2023, Veitch2012_NegativeQuasiProbResource,HowardCampbell2017_RobustnessMagic_PRL,BravyiGosset2016_ImprovedCliffordT_PRL,BravyiSmithSmolin2016_StabilizerExtent_PRX}. However, most of these measures are computationally demanding for many-body states. Recently, the stabilizer Rényi entropy (SRE) has been introduced~\cite{Hamma_PRL_2022,Haug_Monotones2023,Hamma_Renyi_PRA2024,Dowling_MagicHeisenbergPicture_2025} as a measure of the spread of a quantum state over the Pauli basis,
\begin{eqnarray}
\mathcal M_\alpha\left(\lvert\psi\rangle\right)=\frac{1}{1-\alpha}\log_2\left(\,\sum_{ P\in\mathcal P_L}\frac{\lvert\langle\psi\lvert P\rvert\psi\rangle\rvert^{2\alpha}}{2^L}\right),
\end{eqnarray}
where $\mathcal P_L=\left\{\prod_{j=1}^L P_j ,\middle| P_j\in{\sigma_j^s}_{s=0,x,y,z}\right\}$ denotes the set of $L$-site Pauli strings. As a main advantage, it remains efficiently computable by techniques such as matrix product states~\cite{Tarabunga_MPSMagic_2025,Piroli_MPSMagic_2024,Collura_MagicSampling_PRL_2023,Frau_MPS_2024} or via Metropolis-Hastings algorithms~\cite{Tarabunga_Metropolis_2023,Dora_NonHermMagic_2025,Liu_Metropolis_QMagic_2025} and is also experimentally accessible~\cite{Oliviero_npqinf_2022,SalesRodriguez_MSD_2025,Smith_RydbergMagic_2024}.

Understanding how non-stabilizerness is generated and spread under coherent evolution is central to assessing the emergence of quantum computational advantages in physical systems. Existing studies have mostly focused on time-independent Hamiltonians~\cite{Tirrito_AnticoncentrationNonstabilizerness_2025,Falcao_MBLNonstabilizerness_2025}, in non-Hermitian or monitored evolutions~\cite{Bejan_Monitor_2024,Pablo_NonHerm_2025,TirritoNonHermFreeFermion_2025,Fux_DisentanglingMagic_2025,Zarand_QMWalk_2025}, relaxation dynamics after quantum quenches \cite{Rattacaso_PRA_2023,HammaNonInt2025}, quantum walks~\cite{Moca_QMWalk_2025,Huang_QMWalk_2025} and quantum transport~\cite{Tirrito_UniversalSpreadingTransport_2025}, or spreading in quantum circuits~\cite{zhang2024MagicDynamicsRandomCircuit,Turkeshi2025_QuditMagicDynamics,Aditya_QuantumResourcesRandomCircuits_2025,Bejan_MagicSpreadingCliffordDynamics_2025,Aditya_MpembaComplexity_2025,Maity_LocalSREBrickworkClifford_2026}. By contrast, the buildup of magic under explicitly time-dependent driving, especially across quantum phase transitions (QPTs), remains unaddressed, as previous studies have focused solely on static ground-state properties ~\cite{Hamma_TFIM_2022,Tarabunga_IsingCrit_2024}. A key question is how magic dynamics inherits the universal signatures of defect generation across QPTs, such as those captured by the Kibble-Zurek mechanism (KZM)~\cite{Delcampo_Zurek2014,Kibble76a,Kibble76b,Zurek85,Zurek96c,Dziarmaga2005Dynamics}. The KZM provides a universal framework for defect generation for slow driving across continuous QPTs at a constant rate $1/\tau_Q$. It relies on the assumption that adiabaticity breaks down at the freeze-out time scale $\hat t=\left(\tau_0\tau^{z\nu}_Q\right)^{\frac{1}{z\nu+1}}$ setting the typical size of frozen domains, i.e., the freeze-out length scale $\hat\xi=\xi_0(\tau_Q/\tau_0)^{\frac{\nu}{\nu z+1}}$ with $z$ and $\nu$ denoting the equilibrium critical exponents. This immediately implies that the defect number varies as 
 $n \sim \tau^{-\frac{d\nu}{z\nu+1}}_Q$, in $d$ dimensions for point-like defects. Such scaling laws also have a bearing on several entanglement measures \cite{Cherng2006,Cincio_2007,Amico2008,Canovi2014}.  

Quantum many-body systems are generally formulated in terms of local interactions in real space; yet, their low-energy subspace typically exhibits universal signatures that govern the slow dynamics of QPTs, for which an effective momentum-space representation constitutes a powerful tool.
Near criticality, many paradigmatic models reduce to independent quasiparticle modes exactly solvable under linear ramps~\cite{Suzuki2012Quantum,Sachdev2011Quantum,Kitaev_2006}, distilling the universal features of excitation scaling while discarding microscopic details. Momentum space thus offers a natural setting to probe the universal dynamics of non-stabilizerness in nonequilibrium processes.

In this Letter, working in the momentum-space representation naturally suited to the universal slow dynamics near criticality, we establish a direct connection between the universal scaling of defects and the dynamics of magic, the latter captured by the stabilizer R\'enyi entropies and the full statistics of Pauli-spectrum values. Strikingly, we find that the distribution of non-stabilizerness across momentum modes follows a universal lognormal form, while the corresponding cumulants and the SREs relative to the final ground state obey the same universal scaling laws as the defect density. Our predictions are corroborated in the transverse-field Ising and long-range Kitaev models, thereby covering paradigmatic settings in which momentum-space methods reveal universal signatures of QPT dynamics and establishing direct relevance for both fundamental investigations and emerging quantum technologies.

\emph{Non-stabilizerness dynamics in momentum space.---} 
Without loss of generality, we consider homogeneous one-dimensional systems that admit a momentum-space representation,
\begin{eqnarray}\label{eq: Ham_gen}
H&&=\sum_k\hat\psi^\dagger_k\left[(g-\cos\varphi_k)\tau^z\!+\!\sin\theta_k\tau^x\right]\hat\psi_k\nonumber\\
&&=
\sum_{k>0}\hat\psi^\dagger_kH_k\hat\psi_k\!=\!\sum_k\!\epsilon_k\hat\gamma^\dagger_k\hat\gamma_k,\quad
\end{eqnarray}
with $\hat\psi_k=(c_k,c^\dagger_{-k})^T$ containing the fermionic creation and annihilation operators. 
The Bogoliubov operators $\hat c_k=-\cos(\Theta_k/2)\hat \gamma_k+\sin(\Theta_k/2)\hat \gamma^\dagger_{-k}$ diagonalize the Hamiltonians $H_k$ with $\sin\Theta_k=\sin\theta_k/\epsilon_k(g)$. The quasiparticle energies $\epsilon_k(g)=\sqrt{(g-\cos\varphi_k)^2+\sin^2\theta_k}$ close at the critical points $g_c=\pm\cos\varphi_0$, and encode the critical exponents via $\epsilon_k(g_c)\sim k^z$ and $\epsilon_0(g)\sim |g-g_c|^{z\nu}$. These exponents are directly related to the small-momentum behavior $\theta_{k\ll1}\approx C_\beta k^{\beta-1}$ and $\varphi_{k\ll1}\sim k^{\gamma-1}$, yielding $z=\mathrm{min}\{\gamma,\beta\}-1$ and $z\nu=1$.
When the system is slowly driven across the critical point, diabatic excitations are generated. Their number is quantified by the defect operator $\hat N=\sum_{k>0}\hat\gamma^\dagger_k\hat\gamma_k$, with excitation probabilities $p_k=\langle\hat\gamma^\dagger_k\hat\gamma_k\rangle$. For slow ramps, these probabilities are well approximated by Landau–Zener (LZ) transitions,
$p_k\approx e^{-2\pi \tau_Q C_\beta k^{2(\beta-1)}}$,
which leads to a universal power-law scaling of the total number of defects,
$\langle\hat N\rangle=\sum_{k>0}p_k\propto \tau_Q^{-\frac{1}{2(\beta-1)}}$ for $\tau_Q\gg1$, in agreement with the KZM for $\beta\leq\mathrm{min}\{\gamma,2\},\,\gamma<2$ and $\gamma>2$~\cite{Zurek2005Dynamics,Dziarmaga2005Dynamics,Polkovnikov2005,DamskiZurek06,Damski05}.

To capture non-stabilizerness dynamics during such ramps, knowledge of the excitation probabilities alone is insufficient. A refined analysis of the full mode-resolved amplitudes, including both ground- and excited-state components and their relative phases, is required. For a linear ramp $g(t)=-g_0\,t/\tau_Q,\,t\in[-\tau_Q,0]$, with $g_0\gg 1$, the time-evolved state at $t=0$ is obtained from the Schr\"odinger equation
$i\partial_t\left[u_k(t),v_k(t)\right]^T=H_k\left[u_k(t),v_k(t)\right]^T$, $u_k(-\tau_Q)=0,\,v_k(-\tau_Q)=1$. Within the spirit of KZM, the ground- and excited-state amplitudes take the form of 
$v_k\approx \sqrt{1-p_k}\sin\frac{\Theta_k}{2}+e^{i\Phi_k}\sqrt{p_k}\cos\frac{\Theta_k}{2}
    ,\,u_k\approx -\sqrt{1-p_k}\cos\frac{\Theta_k}{2}+\sqrt{p_k}e^{i\Phi_k}\sin\frac{\Theta_k}{2}$ with $p_k$ given by the LZ formula or in a more general setting by the functional form $p_k=p\left(k\tau^{\frac{1}{2(\beta-1)}}_Q\right)$~\cite{Polkovnikov2005}. Here, the relative phase also plays an important role, and for slow driving, is predominantly set by the dynamical contribution accumulated after the impulse regime, $\Phi_k\approx 2\tau_Q$,
to leading order~\cite{Nowak2021,Cincio_2007} (see also~\cite{supp} for further details).

We investigate the nonequilibrium generation of non-stabilizerness in momentum space via the Pauli-spectrum statistics and the SRE, primarily at the end of the quench ($t=0$), while also characterizing the near-critical real-time dynamics around $t_c$. The former is the set of expectation values of the time-evolved state $\lvert\Psi(0)\rangle=\prod_{k>0}\left(u_k\lvert1,1\rangle_{k,-k}+v_k\lvert0,0\rangle_{k,-k}\right)$,

\begin{eqnarray}
\mathrm{spec}\left(\lvert\Psi(0)\rangle\right)=\left\{\lvert\langle\Psi(0)\lvert\Sigma\rvert\Psi(0)\rangle\rvert,\,\Sigma\in\Sigma_L\right\},
\end{eqnarray}
where $\Sigma_L=\left\{\prod_{k>0}\Sigma_k\right\},\,\Sigma_k=\{\sigma_k^s\}_{s=0,x,y,z}$ is the set of the momentum space Pauli string operators. From the possible $16$ configurations for a given  mode $k$, only $8$ give non-zero results,
    $\langle\psi_k\lvert \sigma^0_k\sigma^0_{-k}\rvert\psi_k\rangle=\langle\psi_k\lvert \sigma^z_k\sigma^z_{-k}\rvert\psi_k\rangle$, 
    $\langle\psi_k\lvert \sigma^z_k\sigma^0_{-k}\rvert\psi_k\rangle=\langle\psi_k\lvert \sigma^0_k\sigma^z_{-k}\rvert\psi_k\rangle$, 
    $\langle\psi_k\lvert \sigma^x_k\sigma^y_{-k}\rvert\psi_k\rangle=\langle\psi_k\lvert \sigma^y_k\sigma^x_{-k}\rvert\psi_k\rangle$, $\langle\psi_k\lvert \sigma^x_k\sigma^x_{-k}\rvert\psi_k\rangle=-\langle\psi_k\lvert \sigma^y_k\sigma^y_{-k}\rvert\psi_k\rangle$ (for their general form see~\cite{supp}).
 Although the momentum-space ground state at $g=0$ is not a stabilizer state~\cite{DoraMomSpaceMagic_2025}, it can be taken, without loss of generality, as a reference for studying universal features of non-stabilizerness generation and its relation to defect production. From another perspective, by carefully tuning the ramp and system parameters, one can engineer the final state's degree of non-stabilizerness to lie within an intermediate regime. If it is too low, the system can be efficiently simulated classically; if it is too high, the state behaves essentially randomly under measurement-based protocols, offering no computational advantage. This behavior aligns naturally with the finite ground-state magic, providing a clear reference for systematic control of engineered quantum resources. To this end, we consider the difference between the SREs and their final ground-state values. As shown in~\cite{supp}, this relative SRE takes the form
\begin{eqnarray}\label{eq: DM_alpha}
    \Delta\mathcal M_\alpha&=&\frac{1}{1-\alpha}\log_2\left(\,\prod_{k>0}\frac{\sum_{\Sigma_k}\lvert\langle\psi_k\lvert\Sigma_k\rvert\psi_k\rangle\rvert^{2\alpha}}{\sum_{\Sigma_k}\lvert\langle\mathrm{GS}_k\lvert\Sigma_k\rvert\mathrm{GS}_k\rangle\rvert^{2\alpha}}\right)\nonumber\\
    &\approx&\frac{L}{2\pi(1-\alpha)}I_{\{p(k)\}}\left(2\alpha,\tau_Q\right)\tau^{-\frac{1}{2(\beta-1)}}_Q,
\end{eqnarray}
with $\lvert\mathrm{GS}_k\rangle=(-\cos\Theta_k/2,\sin \Theta_k/2)^T$ denoting the ground state of the $k$-th mode. Here, $I_{\{p(k)\}}\left(2\alpha,\tau_Q\right)$ is an oscillatory bounded function dependent also on the functional form of the excitation probabilities, $p(k)$, with negligible impact on the overall scaling~\cite{supp}. 
 Consequently, although the absolute amount of magic depends on the representation and may differ in real space compared to momentum space, the relative deviations from the final ground-state SRE and their scaling with respect to $\tau_Q$
 remain robust, justifying a common universal description linking non-stabilizerness generation and critical dynamics across QPTs. Notably, the universal relation between non-stabilizerness generation and defect production persists even for $\alpha<1$, where the SRE is not a magic monotone.~\cite{Haug_Monotones2023,Hamma_Renyi_PRA2024}

Next, we analyze the distribution of the Pauli spectrum values. To access large system sizes while avoiding exponentially diverging quantities, we adopt a logarithmic representation of the Pauli spectrum, providing a more meaningful and tractable analysis. The corresponding statistics is constructed from the histogram counts of the occurrences of the Pauli spectrum values. In the thermodynamic limit, this is equivalent to the sum of independent four-state uniform discrete random variables,
\begin{eqnarray}
    P_\mathrm{log}\left(x\right)=8^{-L/2}\sum_{\Sigma\in\Sigma_L}\delta\left[x-\log\left(\left\lvert\left\langle\Psi\left\lvert\Sigma\right\rvert\Psi\right\rangle\right\rvert\right)\right],
\end{eqnarray}
where the normalization excludes cases with zero Pauli spectrum values. Owing to the imprinted independence of the 
$k$-modes, the limiting distribution is Gaussian according to Lindeberg's theorem, while broader statistical features are captured by the cumulant generating function. While naive histogram evaluation would grow as $4^L$, we show in~\cite{supp} that the underlying analytical structure of the model enables a compact algorithm, 
simplifying earlier techniques based on MPS and Monte Carlo methods.
In analogy with the SRE analysis, we quantify deviations from the final ground-state Pauli spectrum by the difference of the cumulant generating functions of the logarithmic Pauli spectrum. As shown in~\cite{supp}, it can be expressed similarly to the relative SRE in Eq.~\eqref{eq: DM_alpha},
\begin{eqnarray}\label{eq: cum_gen}
    \Delta \log\tilde P_\mathrm{log}(\theta)&=&\Delta\log\mathbb E\left[\left\lvert\left\langle\Psi\left\lvert\Sigma\right\rvert\Psi\right\rangle\right\rvert^{i\theta}\right]\\
    &\approx&\frac{L}{2\pi}I_{\{p(k)\}}(i\theta,\tau_Q)\tau^{-\frac{1}{2(\beta-1)}}_Q,\nonumber
\end{eqnarray}
following the same universal power-law with the same integral function. Since $I_{\{p(k)\}}(i\theta,\tau_Q)$ only enters in an oscillatory way~\cite{supp}, Eq.~\eqref{eq: cum_gen} immediately implies that the cumulants of the logarithmic Pauli spectrum are proportional to the mean, thus following the same power-law
\begin{eqnarray}
\kappa^{(\mathrm{log})}_q\propto\tau^{-\frac{1}{2(\beta-1)}}_Q.
\end{eqnarray}
 As a result, the histogram of the Pauli spectrum exhibits a universal lognormal distribution, fully characterized by $\kappa^{(\mathrm{log})}_{1,2}$,
\begin{eqnarray}
    P(x)=\sum_{\Sigma\in\Sigma_L}\frac{\delta\left[x-\lvert\langle\Psi\lvert\Sigma\rvert\Psi\rangle\rvert\right]}{4^L}\rightarrow \frac{e^{-\frac{\left(\log x-\kappa^{(\log)}_1\right)^2}{2\kappa^{(\log)}_2}}}{x\sqrt{2\pi\kappa^{(\log)}_2}}.\quad  
\end{eqnarray}
This behavior, rooted in the underlying integrability, is fundamentally distinct from typical many-body states such as Haar-random vectors, eigenstates of chaotic Hamiltonians, and generic non-integrable systems~\cite{Turkeshi_PRL_2025,NahumVijayHaah2018_OperatorSpreading_PRX}.
\begin{figure}
    \includegraphics[width=.99\linewidth]{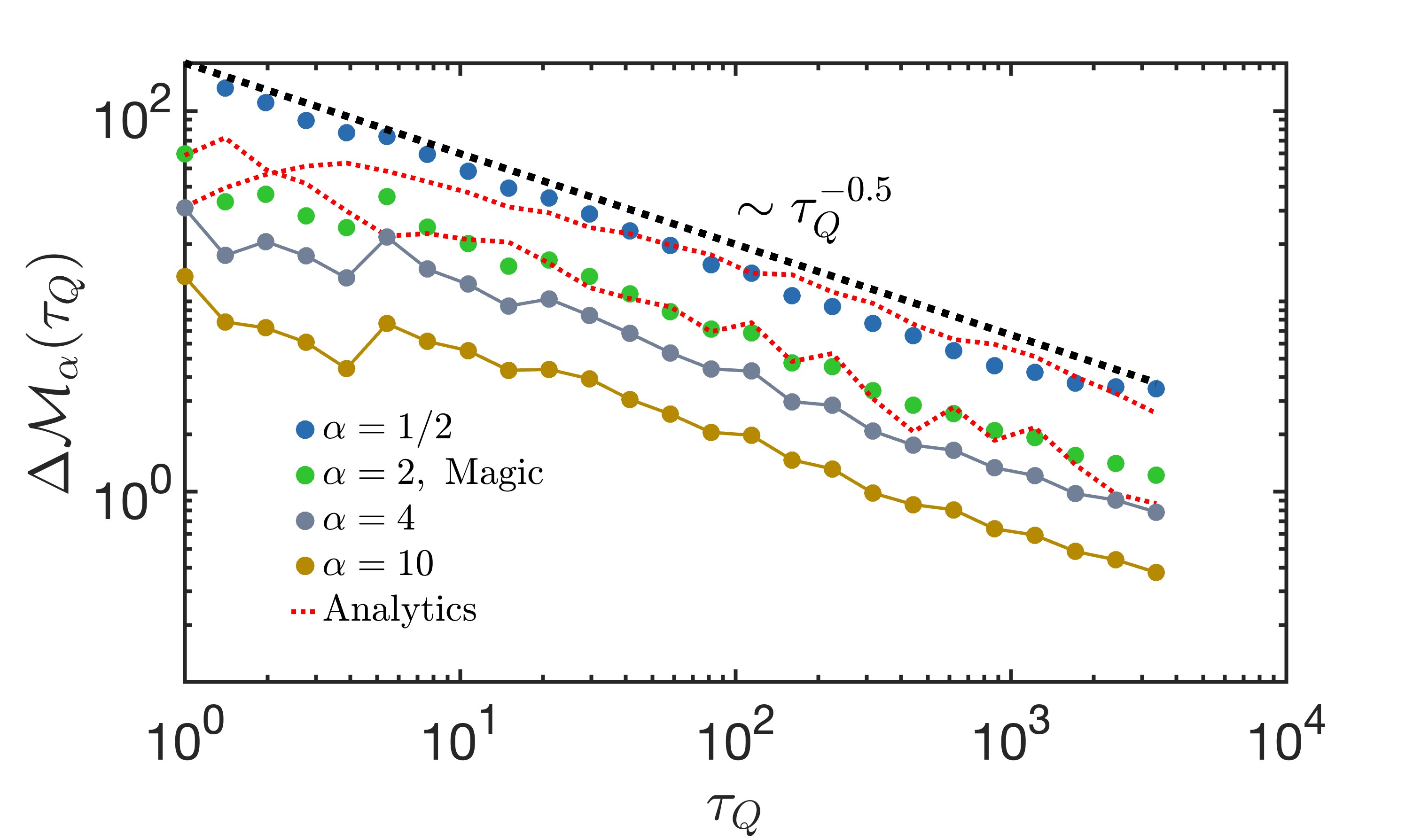}\\
    \caption{Stabilizer R\'enyi entropies versus $\tau_Q$ for various $\alpha$ in the TFIM. All curves show the predicted universal power-law decay with weak superimposed oscillations, captured by the analytical approximations (red dashed, $L=1600$).}
    \label{fig:TFIM_M_alpha}
\end{figure}
\begin{figure}[t]
    \includegraphics[width=.99\linewidth]{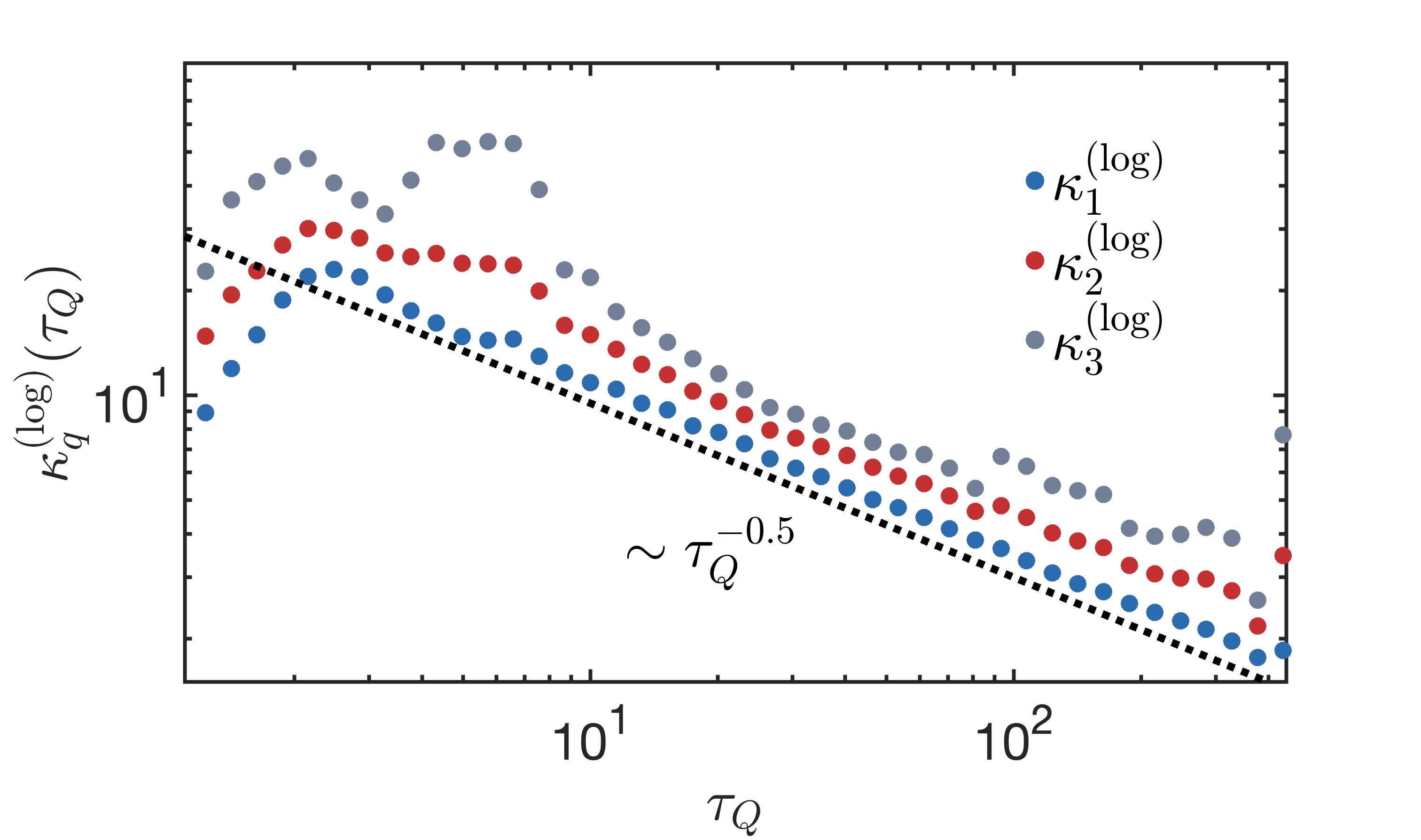}\\
    \caption{First three cumulants of the logarithmic Pauli spectrum for the TFIM, showing precise agreement with the predicted KZ power-law $(L=200)$.}
    \label{fig:TFIM_cumulants}
\end{figure}
The leading power laws are independent of $\Theta_k$, which enters only through oscillatory prefactors. This indicates that the same scaling is likely to hold more broadly whenever $p_k$ admits the universal form $p_k=p(k\tau_Q^{\delta})$~\cite{Polkovnikov2005}, implying $n\sim\tau_Q^{-\delta}$ and
\begin{eqnarray}
\Delta\mathcal M_\alpha\propto \tau^{-\delta}_Q,\quad \kappa^{(\mathrm{log})}_q\propto \tau^{-\delta}_Q.
\end{eqnarray}
Further analytical details are shown in~\cite{supp}, with the general framework illustrated in the specific models below.

\emph{Transverse-field Ising model (TFIM).---}
First, we demonstrate the universal magic generation in the TFIM, described
by the Hamiltonian~\cite{Suzuki2012Quantum}
\begin{equation}
	\label{eq:H_TFIM}
	\hat{H}(t) = - J \sum_{j=1}^L \left[ \hat{\sigma}^z_j \hat{\sigma}^z_{j+1} + g(t) \hat{\sigma}_j^x \right],
\end{equation}
an ideal testbed for studying equilibrium and non-equilibrium properties of QPTs~\cite{Zurek2005Dynamics, Dziarmaga2005Dynamics,delCampo2018Universal, Bando20, King22, Balducci2023Large}.
Choosing ferromagnetic couplings, $J\equiv 1$, the critical points are given by $g_c=\pm1$~\cite{Sachdev2011Quantum}. 
Mapping the Hamiltonian to free fermions by the Jordan-Wigner transformation by $\sigma^x_j=1-2c^\dagger_jc_j,\,\sigma^z_j=-(c_j+c^\dagger_j)\prod_{m<j}(1-2c^\dagger_mc_m)$ and further decomposing in Fourier space, $c_j=e^{-i\pi/4}\sum_kc_ke^{ikj}/\sqrt L$, one arrives at the sum of independent two-level systems (TLSs) given by the Hamiltonian of the general form in Eq.~\eqref{eq: Ham_gen} with $\varphi_k=\theta_k=k$ and $z=\nu=1$,
 with momenta $k=(2m+1)\pi/L,\,m=0,\dots,L-1$. Accordingly, the relative SRE with $p(k)=e^{-2\pi k^2}$ given by the LZ formula reads
\begin{eqnarray}\label{eq: M_alpha_TFIM}
    \Delta\mathcal M_\alpha\approx \frac{L}{2\pi}I_{\{p(k)\}}(\alpha,\tau_Q)\tau^{-1/2}_Q.
\end{eqnarray} 

\begin{figure}
    \includegraphics[width=.99\linewidth]{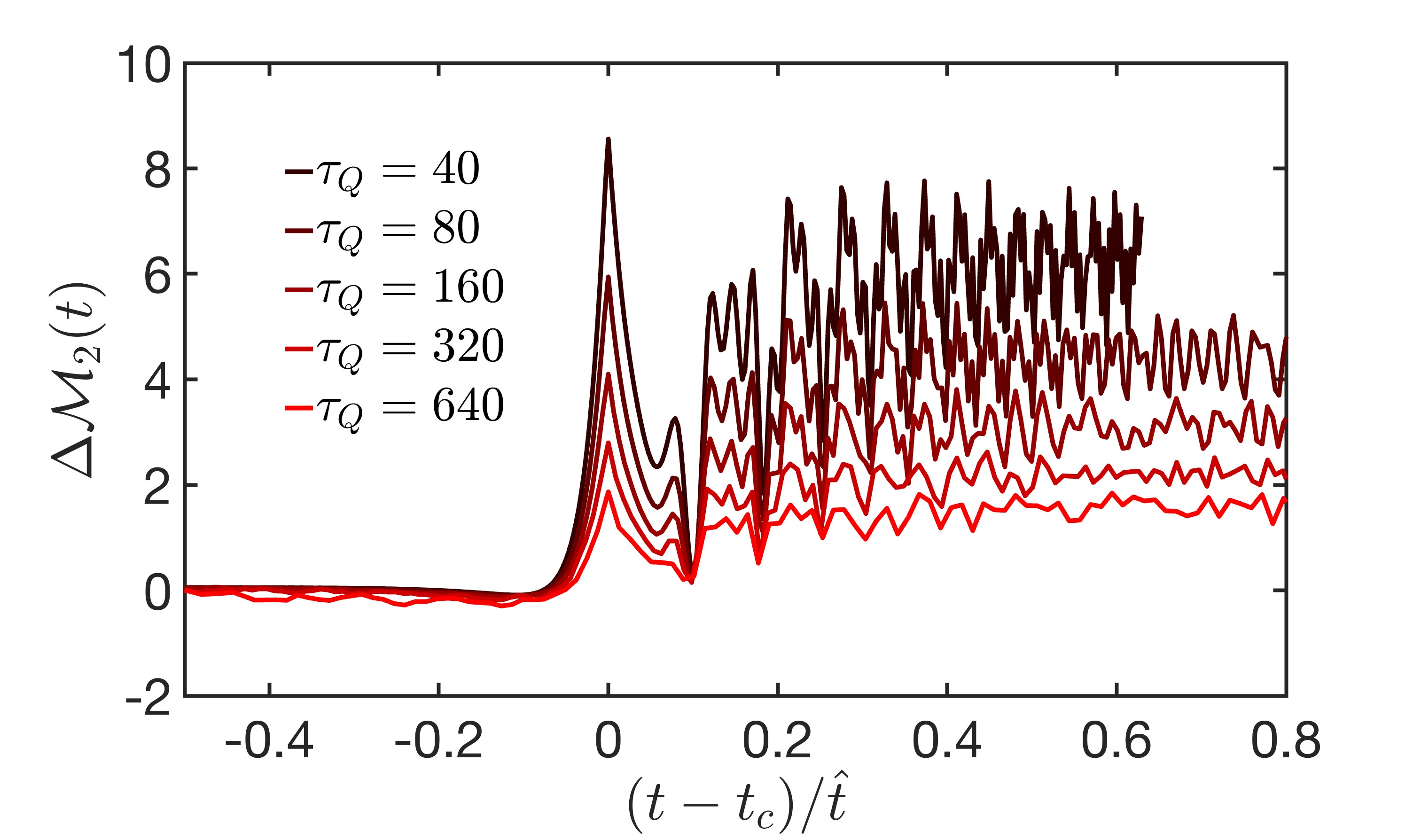}\\
    \caption{Universal time evolution of quantum magic relative to the instantaneous ground state for different driving rates in the TFIM. The time of evolution is measured relative to the instant $t_c$ at which the critical point is reached and scaled by the freeze-out time $\hat{t}$. Near the critical point, the curves exhibit a sudden increase and collapse onto a universal scaling form, followed by an oscillatory intermediate regime ($L=1600$).}
    \label{fig:M_alpha_t}
\end{figure}

As shown in Fig.~\ref{fig:TFIM_M_alpha}, the additional oscillatory behavior with the non-universal $\tau_Q$ dependence provides a negligible correction compared to the universal KZ-type decay, while oscillations of the quantum magic for $\alpha=2$ and of $\Delta\mathcal M_{1/2}$ are accurately captured by the large $\tau_Q$ approximations of $I_{\{p(k)\}}(\alpha,\tau_Q)$~\cite{supp}.
In full agreement with the general framework, the cumulants of the logarithmic Pauli spectrum exhibit a universal decay, $\kappa^{(\log)}_q\propto\tau^{-1/2}_Q$, as shown in Fig.~\ref{fig:TFIM_cumulants}.
Last, we examine the real-time evolution of the SRE relative to the instantaneous ground-state values. As shown in Fig.~\ref{fig:M_alpha_t}, the quantum magic adiabatically tracks the ground-state behavior in the paramagnetic phase ($g>1$), while exhibiting a sharp increase upon crossing the critical point.
Remarkably, the SRE dynamics reproduces even the KZM near-critical universality, with curves for different $\tau_Q$ collapsing on top of each other around $t_c$ when plotted as a function of $(t-t_c)/\hat t$.
This is followed by a nonuniversal, strongly oscillatory regime in the ferromagnetic phase $|g|<1$ relaxing toward Eq.~\eqref{eq: M_alpha_TFIM}.

\begin{figure}
    \includegraphics[width=.99\linewidth]{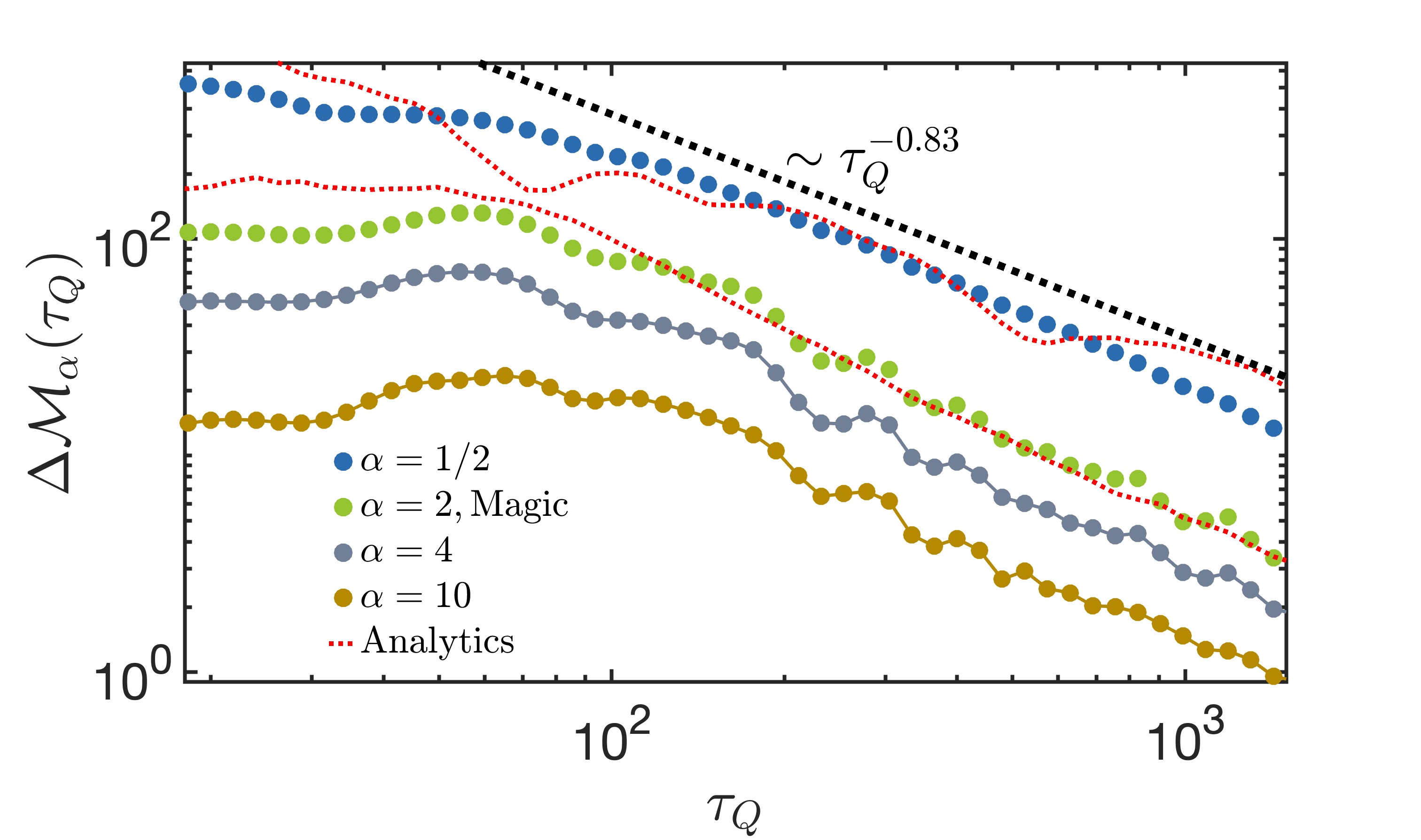}\\
    \caption{Quantum magic, $\alpha=1/2$, and higher order SREs relative to the final ground state in the LRKM following precisely the predicted dynamical scaling laws for $\gamma=1.4,\,\beta=1.6$, in close agreement with the analytical approximations for $\alpha=1/2,\,\alpha=2$ ($L=1000$).}
    \label{fig:LRKM_M_alpha}
\end{figure}

\begin{figure}
    \includegraphics[width=.99\linewidth]{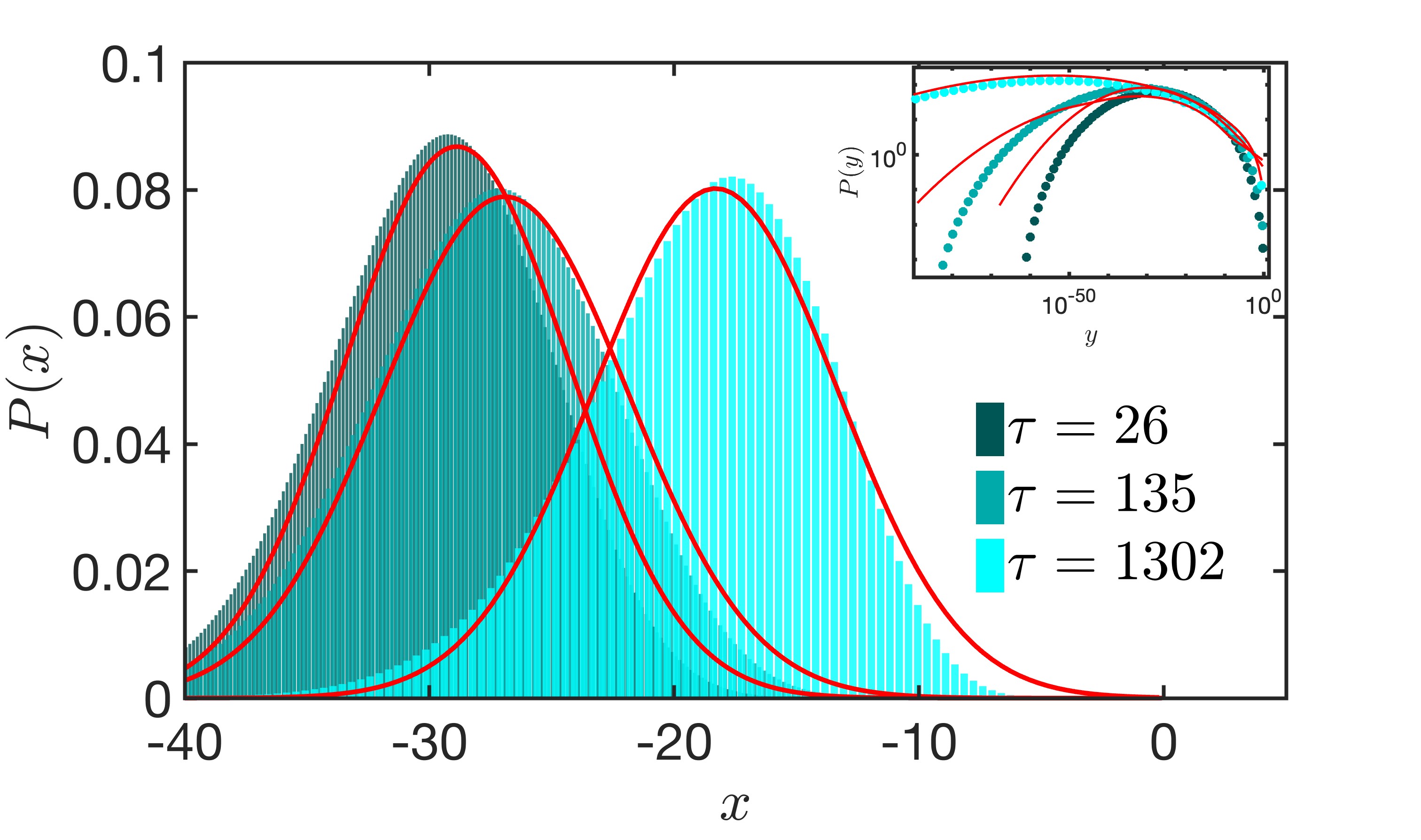}\\
    \caption{Pauli spectrum statistics in the LRKM ($\gamma=1.4,\,\beta=1.6$).  The approximately Gaussian statistics of the logarithmic Pauli spectrum converges to the ground-state distribution with increasing driving times. Inset: Pauli spectrum distributions following the corresponding lognormal distributions.}
    \label{fig:LRKM_cum}
\end{figure}

\emph{Long-range Kitaev models (LRKMs).---}
Finally, we demonstrate the validity of our results by considering LRKMs describing spinless fermions with hopping and $p$-wave pairing interactions on a one-dimensional lattice, which provides 
an explicit realization of the general form Eq.~\eqref{eq: Ham_gen},
\begin{eqnarray}
\hat H\!=\!-\sum_i\sum_{r>0}\!\!\left(j_{r,\gamma} c_i^\dagger c_{i+r}\!+\!\Delta_{r,\beta} c_i^\dagger c_{i+r}^\dagger\right)\!+\!\mu\, c_i^\dagger c_i
+\mathrm{h.c.}\quad\,\,\,\,\,\,
\end{eqnarray}
The hopping and pairing amplitudes decay algebraically, 
$j_{r,\gamma}=J/N_\gamma\, r^{-\gamma}$ and $\Delta_{r,\beta}=d/N_\beta\, r^{-\beta}$, 
with exponents $\gamma,\beta>1$ and normalization factors $N_{\gamma,\beta}=2\sum_{r=1}^{N/2}r^{-\gamma,\,-\beta}$ ensuring extensivity and with $J=d=1$ taken for simplicity.
With these exponents, the LRKM exhibits a second-order QPT at $\mu_c=2$~\cite{Kitaev_Majorona_LR_2001}, which is robust against variations of the long-range exponents $\gamma,\beta>1$~\cite{Vodola_LRK, Dutta_LRK, LongRangePowerLawSC_Delgado_2017, LRK_powerlawDelAnna2017, Defenu_LRKM}. The resulting excitation density for slow drivings is fully governed by the pairing term, $\langle\hat N\rangle\propto \tau^{-\frac{1}{2(\beta-1)}}_Q,\,\beta <2,\,$ and $\langle\hat N\rangle\propto \tau^{-1/2}_Q,\,\beta >2,\,$ which violate the KZ scaling prediction in the dynamical scaling regime, for $\gamma<\beta,\,\gamma<2$. Our results are in perfect agreement with the general predictions, Eq.~\eqref{eq: DM_alpha}, as shown in Fig.~\ref{fig:LRKM_M_alpha} for the SRE converging to the final GS value as $\tau^{-0.83}_Q$ with $\gamma=1.4,\,\beta=1.6$. Additionally, Fig.~\ref{fig:LRKM_cum} demonstrates the Gaussian shape of the logarithmic Pauli spectrum histograms with the inset showing the matching with the corresponding lognormal statistics of the Pauli spectrum itself. The universal scaling of the cumulants of the logarithmic Pauli spectrum is shown in~\cite{supp} together with the results in the different power-law regimes of the LRKMs and the universal near-critical dynamics of the SRE.

\emph{Conclusions.---}
We have shown that the non-stabilizerness generated during slow quenches across quantum phase transitions exhibits universal scaling behavior directly tied to defect formation. In particular, the stabilizer Rényi entropies and the cumulants of the Pauli spectrum display the same power-law dependence on the driving rate, while the full distribution of non-stabilizerness converges to a universal lognormal form.
Importantly, slow driving keeps the excess non-stabilizerness relative to the final ground state controlled, demonstrating that non-stabilizerness is not merely a resource to be generated, but also a quantity that can be systematically tuned in nonequilibrium processes.
Validated in the TFIM and LRKMs, these findings are likely to extend to a broader class of systems regardless of the validity of the KZM. This establishes a universal connection between critical dynamics and controllable quantum resources, and motivates experimental tests.

{\it Acknowledgements.---}
This project was supported by the Luxembourg National Research Fund under Grants No.
C22/MS/17132060/BeyondKZM, C22/MS/17132054/AQCQNET, and CC25/MS/19559370/FastQOPT.

\bibliography{references}

\newpage
\clearpage

\title{---Supplementary Material---Universal Non-stabilizerness Dynamics Across Quantum Phase Transitions
}
\maketitle
\onecolumngrid
\begin{center}
\textbf{\large Supplemental Material for\\
	    ``Universal Non-stabilizerness Dynamics Across Quantum Phase Transitions''}\\
\vspace{0.5cm}
 Andr\'as Grabarits and Adolfo del Campo 
\end{center}
\renewcommand{\theequation}{S\arabic{equation}}
\renewcommand{\thefigure}{S\arabic{figure}}
\renewcommand{\thetable}{S\arabic{table}}

\setcounter{equation}{0}
\setcounter{figure}{0}
\setcounter{table}{0}
\setcounter{page}{1}

    In this Supplementary Material, further notes and extended results are provided to support and elucidate the findings in the main text. First, we show the details of the calculation of the SRE and the Pauli spectrum in momentum space, including also the general scaling argument for the freeze-out time and the leading order relative phase. Next, we show the derivation of the exact results for the TFIM. Then, we show additional numerical demonstrations of the statistics of the Pauli spectrum by its full distribution and cumulants, both for the LRKMs and the TFIM. Finally, we show the results regarding the R\'enyi index scaling of the SRE for both the LRKMs and TFIM.
\maketitle%

\section{Freeze-out time and relative phases in quantum critical dynamics}

In this section, we provide a compact, approximate analytical treatment of the relative phases that arise during the slow driving of the QPT. Exploiting the adiabatic regime, the finite amplitude of the excited state in each $k$-mode appears after the freeze-out time. For analytical convenience, we restrict the derivations to the general Hamiltonian
\begin{eqnarray}
    H_k=(g-\cos\varphi_k)\sigma^z_k+\sin\theta_k\sigma^x_k,\,\theta_{k\leq 1}\sim k^{\beta-1},\,\varphi_{k\ll 1}\sim k^{\gamma-1}.
\end{eqnarray}
We start with the case of $\beta<\gamma$, for which the corresponding freeze-out time in the $k$-th TLS is given by the condition of the adiabatic breakdown, i.e., the momentum space refined boundary of the impulse regime for the low energy spectrum, given by $k\lesssim \tau^{-\frac{1}{2(\beta-1)}}_Q$
\begin{eqnarray}\label{eq: t_hat}
&&\epsilon_k=\sqrt{(t/\tau_Q)^2+k^{2(\beta-1)}},\quad\mathrm d\epsilon_k/\mathrm dt=\frac{t/\tau_Q}{\tau_Q\sqrt{(t/\tau_Q)^2+k^{2(\beta-1)}}},\\
    &&\epsilon^2_k\sim\mathrm d\epsilon_k/\mathrm dt\rightarrow \tau_Q\left((t_k/\tau_Q)^2+k^{2(\beta-1)}\right)^{3/2}=t_k/\tau_Q\Rightarrow t_k\sim\sqrt{\tau_Q},
\end{eqnarray}
as $(t_k/\tau_Q)^2\sim 1/\tau_Q$ and $k^{2(\beta-1)}\lesssim1/\tau_Q$ that match the overall scaling. For the dynamical scaling regime with $\gamma<\beta$ the leading order small momentum expansion is given by
\begin{eqnarray}\label{eq: t_hat_LRKM}
    &&\epsilon_k=\sqrt{\left(t/\tau_Q-k^{\gamma-1}\right)^2+k^{2(\mathrm{min}\{2,\beta\}-1)}},\,\mathrm d\epsilon_k/\mathrm dt=\frac{t/\tau_Q-k^{\gamma-1}}{\tau_Q\sqrt{\left(t/\tau_Q-k^{\gamma-1}\right)^2+k^{2(\mathrm{min}\{2,\beta\}-1)}}},\\
    &&\epsilon^2_k\sim\mathrm d\epsilon_k/\mathrm dt\Rightarrow \left(t/\tau_Q-k^{\gamma-1}\right)^3\sim\frac{t/\tau_Q-k^{\gamma-1}}{\tau_Q}\Rightarrow t_k/\tau_Q\sim \tau^{-1/2}_Q+k^{\gamma-1}\Rightarrow t_k\sim\sqrt{\tau_Q}+k^{\gamma-1}\tau_Q,\nonumber\\
\end{eqnarray}
where we used that $\sqrt{\tau_Q}\lesssim k^{-(\beta-1)},\,k^{-(\beta-1)}\gg k^{-(\gamma-1)}$ and so one remains with two momentum regimes. However, the breakdown of adiabaticity is governed by the lowest-lying modes $\sqrt{\tau_Q}\lesssim k^{-(\gamma-1)}$, implying $\tau_Qk^{\gamma-1}\lesssim \sqrt{\tau_Q}$ and a freeze-out time-scale $\hat t\sim\sqrt{\tau_Q}$.
 These exact scaling arguments can naturally be extended to arbitrary critical systems, where for low-lying energy states the freeze-out time is given by the conventional scaling $\hat t\propto\tau^{\frac{z\nu}{z\nu+1}}_Q$ when the KZM holds, while its effective version can be constructed via the defect power-law scaling in cases when the KZM is violated,
\begin{eqnarray}
\langle N\rangle\propto \tau^{-\delta}_Q\propto (\hat t/\tau_Q)^\nu\Rightarrow\hat t\propto\tau^{1-\delta/\nu}_Q\ll\tau_Q.
\end{eqnarray}
As a result, the low-energy relative phases are determined by the dynamical one, which are given in the leading order by
\begin{eqnarray}
    \Phi_k\approx2\int_{\hat t}^{\tau_Q}\mathrm dt\sqrt{(t/\tau_Q)^2+k^{2(\beta-1)}}=2\tau_Q\left[\frac{t}{\tau_Q}\sqrt{\frac{t^2}{\tau^2_Q}+k^{2(\beta-1)}}+k^{2(\beta-1)}\log\left[\frac{t}{\tau_Q}+\sqrt{\frac{t^2}{\tau^2_Q}+k^{2(\beta-1)}}\right]\right]\Bigg\vert_{\hat t}^{\tau_Q}.
\end{eqnarray}
This expression is dominated by $\tau_Q\gg\hat t$, thus in the leading order it becomes
\begin{eqnarray}
    \Phi_k\approx2\tau_Q(1+O(k^{2(\beta-1)}\log k)\approx2\tau_Q.
\end{eqnarray}
In the dynamical scaling regime, one has simpler relations,
\begin{eqnarray}
    \Phi_k\approx2\int_{\hat t}^{\tau_Q}\mathrm dt\left[\frac{t}{\tau_Q}-k^{\gamma-1}\right]=\tau_Q(1+O(k^{\gamma-1}))\approx 2\tau_Q,
\end{eqnarray}
justifying the main statement.

\section{Details of the calculations for the stabilizer R\'enyi entropy}
First, we show the detailed steps to obtain the final scaling for most stabilizer R\'enyi entropies, reported in Eq.~\eqref{eq: DeltaSRE} of the main text. The matrix elements of the Pauli spectrum for the time-evolved states
\begin{eqnarray}
    &&\langle\psi_k\lvert \sigma^0_k\sigma^0_{-k}\rvert\psi_k\rangle=\langle\psi_k\lvert \sigma^z_k\sigma^z_{-k}\rvert\psi_k\rangle=1,\\
    &&\langle\psi_k\lvert \sigma^z_k\sigma^0_{-k}\rvert\psi_k\rangle=\langle\psi_k\lvert \sigma^0_k\sigma^z_{-k}\rvert\psi_k\rangle=\lvert u_k\rvert^2-\rvert v_k\lvert^2=2p_k-\cos\Theta_k,\\
    &&\langle\psi_k\lvert \sigma^x_k\sigma^y_{-k}\rvert\psi_k\rangle=\langle\psi_k\lvert \sigma^y_k\sigma^x_{-k}\rvert\psi_k\rangle=-2\,\mathrm{Im}\{u_k v^*_k\}=-2\sqrt{p_k(1-p_k)}\sin\Phi_k,\\
    &&\langle\psi_k\lvert \sigma^x_k\sigma^x_{-k}\rvert\psi_k\rangle=2\mathrm{Re}\{u_kv^*_k\}=-\langle\psi_k\lvert \sigma^y_k\sigma^y_{-k}\rvert\psi_k\rangle=\sin \Theta_k+2\sqrt{p_k(1-p_k)}\cos\Phi_k,
\end{eqnarray}
and for the ground states of each $k$-th mode
\begin{eqnarray}
&&\lvert\mathrm{GS}(-\tau_Q)\rangle=\prod_{k>0}\lvert\mathrm{GS}_k(-\tau_Q)\rangle,\quad \lvert\mathrm{GS}_k(-\tau_Q)\rangle=\left(-\cos\frac{\Theta_k}{2},\sin\frac{\Theta_k}{2}\right)^T,\\
    &&\langle\mathrm{GS}_k(-\tau_Q)\lvert \sigma^0_k\sigma^0_{-k}\rvert\mathrm{GS}_k(-\tau_Q)\rangle=\langle\mathrm{GS}_k(-\tau_Q)\lvert \sigma^z_k\sigma^z_{-k}\rvert\mathrm{GS}_k(-\tau_Q)\rangle=1,\\
    &&\langle\mathrm{GS}_k(-\tau_Q)\lvert \sigma^z_k\sigma^0_{-k}\rvert\mathrm{GS}_k(-\tau_Q)\rangle=\langle\mathrm{GS}_k(-\tau_Q)\lvert \sigma^0_k\sigma^z_{-k}\rvert\mathrm{GS}_k(-\tau_Q)\rangle=-\cos \Theta_k,\\  
    &&\langle\mathrm{GS}_k(-\tau_Q)\lvert \sigma^x_k\sigma^x_{-k}\rvert\mathrm{GS}_k(-\tau_Q)\rangle=-\langle\mathrm{GS}_k(-\tau_Q)\lvert \sigma^y_k\sigma^y_{-k}\rvert\mathrm{GS}_k(-\tau_Q)\rangle=\sin \Theta_k.
    \end{eqnarray}
For the stabilizer R\'enyi entropy of the time-evolved states, we first compute the argument of the logarithm,
    \begin{eqnarray}\label{eq: Pauli_spec}
&&\sum_{\Sigma\in \Sigma_L}\lvert\langle\Psi\lvert \Sigma\vert\Psi\rangle\rvert^{2\alpha}=\sum_{\Sigma\in\Sigma_L}\prod_{k>0}\left\lvert\langle\psi_k\lvert \Sigma_k\Sigma_{-k}\rvert\psi_k\rangle\right\rvert^{2\alpha}=\prod_{k>0}\sum_{\Sigma_k} \left\lvert\langle\psi_k\lvert \Sigma_k\Sigma_{-k}\rvert\psi_k\rangle\right\rvert^{2\alpha}\\
&&=2^{L/2}\prod_{k>0}\left[1+\left\lvert\cos\Theta_k-2p_k\right\rvert^{2\alpha}+\left\lvert2\sqrt{p_k(1-p_k)}\sin\Phi_k\right\rvert^{2\alpha}+\left\lvert\sin\Theta_k+2\sqrt{p_k(1-p_k)}\cos\Phi_k\right\rvert^{2\alpha}\right].\nonumber
\end{eqnarray}
As the ground state values of the Pauli spectrum are reproduced by simply taking the limit $p_k\rightarrow0$, one obtains for the difference of stabilizer R\'enyi entropies
\begin{eqnarray}\label{eq: DeltaSRE}
    &&\Delta\mathcal M_\alpha=\frac{1}{1-\alpha}\times\\
& & \sum_{k>0}\log_2\!\left[
\frac{1
+(\cos\Theta_k-2p_k)^{2\alpha}
+\left(\sin\Theta_k+2\sqrt{p_k(1-p_k)}\cos(2\tau_Q)\right)^{2\alpha}
+\left(4p_k(1-p_k)\sin^2(2\tau_Q)\right)^{\alpha}
}{
1+(\cos\Theta_k)^{2\alpha}+(\sin\Theta_k)^{2\alpha}
}
\right].\nonumber    
\end{eqnarray}
Using that the $p_k$ probabilities restrict the momenta to $k\lesssim \tau^{-\frac{1}{2(\beta-1)}}_Q$ and that at $g=0$ $\Theta_k=\theta_k\sim k^{\beta-1}$ for small $k$, the second and third terms in the logarithm can be expanded up to leading order as
\begin{eqnarray}
    &&(\cos \Theta_k-2p_k)^{2\alpha}-(\cos \Theta_k)^{2\alpha}=\sum_{n=1}^\infty\binom{2\alpha}{n}(-2p_k)^n\cos^{2\alpha-n}\Theta_k=(1-2p_k)^{2\alpha}-1+O(\tau^{-1/2}_Q),\\
    &&\left(\sin\Theta_k+2\sqrt{p_k(1-p_k)}\cos(2\tau_Q)\right)^{2\alpha}-(\sin \Theta_k)^{2\alpha}=\sum_{n=1}^\infty\binom{2\alpha}{n}2^n p^{n/2}_k(1-p_k)^{n/2}\cos^{n}(2\tau_Q)\sin^{2\alpha-n}\Theta_k\\
    &&\qquad\qquad\qquad\qquad\qquad\qquad\quad\qquad\qquad\qquad\qquad\approx 4^\alpha p^\alpha_k(1-p_k)^{\alpha}\cos^{2\alpha}(2\tau_Q)+O(\tau^{-1/2}_Q).\nonumber
\end{eqnarray}
At the same time, the denominator is also restricted to the low momentum region as $1+(\cos\Theta_k)^{2\alpha}+(\sin\Theta_k)^{2\alpha}=2+O(\tau^{-1/2}_Q)$.
As a result, the leading order expression reads
\begin{eqnarray}\label{eq: Delta_M_alpha}
    \Delta\mathcal M_\alpha&&\approx \frac{1}{1-\alpha}\sum_{k>0}\log_2
\left[
\frac{
1+(1-2p_k)^{2\alpha}
+\bigl(4p_k(1-p_k)\bigr)^\alpha
\left(\sin^{2\alpha}(2\tau_Q)+\cos^{2\alpha}(2\tau_Q)\right)
}{2}
\right]\nonumber\\
    &&\approx\frac{L}{2\pi(1-\alpha)}I_{\{p(k)\}}(2\alpha,\tau_Q)\tau^{-\frac{1}{2(\beta-1)}}_Q,
\end{eqnarray}
where the last term approaches zero at the same speed as the $p_k$ rates vanish, and where the integral function has been defined as
\begin{eqnarray}\label{eq: I_alpha}
I_{\{p(x)\}}(2\alpha,\tau_Q)
=
\int_0^\infty \mathrm{d}x\,
\log_2\!\left[
\frac{
1+(1-2p(x))^{2\alpha}
+\left[4p(x)(1-p(x))\right]^{\alpha}
\left(\sin^{2\alpha}(2\tau_Q)+\cos^{2\alpha}(2\tau_Q)\right)
}{2}
\right].
\end{eqnarray}
In the last step, the universal scaling was assumed for the excitation probabilities $p_k=p(\tau^{\delta}_Q k)$ responsible for defect generation in the low-energy region. For the TFIM and the LRKMs, this takes the form of the LZ transition probabilities, i.e., $p(x)=e^{-2\pi C_\beta x^{2(\beta-1)}}$. The form of the integral function highlights that it depends on the functional form of the low momentum excitation probabilities, $\{p(k)\}$ in the discrete and on $\{p(x)\}$ in the continuum representations.

\section{Details of the derivations for the statistics of the Pauli spectrum}
In this section, we present the details of the calculations that yield the universal scalings of the cumulants of the logarithmic Pauli spectrum. We start from the cumulant generating function of the logarithmic Pauli spectrum, defined as the logarithm of the characteristic function,
\begin{eqnarray}
    P_\mathrm{log}\left(x\right)&=&2^{-3L/2}\sum_{\Sigma\in\Sigma_L}\delta\left[x-\log\left(\left\lvert\left\langle\Psi\left\lvert\Sigma\right\rvert\Psi\right\rangle\right\rvert\right)\right],\\
    \tilde P_\mathrm{log}(\theta)&=&\mathbb E\left[e^{i\theta\log\left(\left\lvert\left\langle\Psi\left\lvert\Sigma\right\rvert\Psi\right\rangle\right\rvert\right)}\right]=\mathbb E\left[\left\lvert\left\langle\Psi\left\lvert\Sigma\right\rvert\Psi\right\rangle\right\rvert^{i\theta}\right]=2^{-3L/2}\sum_{\Sigma\in\Sigma_L}\left\lvert\left\langle\Psi\left\lvert\Sigma\right\rvert\Psi\right\rangle\right\rvert^{i\theta}\\
    &=&2^{-3L/2}\prod_{k>0}\left[1+\left\lvert\cos \Theta_k-2p_k\right\rvert^{i\theta}+\left\lvert2\sqrt{p_k(1-p_k)}\sin\Phi_k\right\rvert^{i\theta}+\left\lvert\sin \Theta_k+2\sqrt{p_k(1-p_k)}\cos\Phi_k\right\rvert^{i\theta}\right],\nonumber
\end{eqnarray}
where we have excluded those taking zero values.
It takes the same form as the stabilizer R\'enyi entropy with $\alpha\equiv i\theta$ apart from the factor $1/(1-\alpha)$. As a result, repeating the same steps as in Eq.~\eqref{eq: Delta_M_alpha} and Eq.~\eqref{eq: I_alpha}, one arrives at the approximate formula for the cumulant generating function relative to the final ground state,
    \begin{eqnarray}\label{eq:logP_log}
    &&\Delta \log\tilde P_\mathrm{log}(\theta;\tau_Q)\approx\frac{L}{2\pi}I_{\{p(k)\}}(i\theta;\tau_Q)\tau^{-\frac{1}{2(\beta-1)}}_Q,
\end{eqnarray}
with the same integral function as in Eq.~\eqref{eq: I_alpha}. As this term only contains $\tau_Q$ via the leading order expression of the phases in an oscillatory way, its derivatives will not change the overall leading order power-law scaling with respect to $\tau_Q$,
\begin{eqnarray}
    \kappa^{(\log)}_q=\frac{L}{2\pi}\tau^{-\frac{1}{2(\beta-1)}}_Q\partial^q_{i\theta} I_{\{p(k)\}}(i\theta;\tau_Q)\big\vert_{\theta=0}\propto \tau^{-\frac{1}{2(\beta-1)}}_Q.
\end{eqnarray}
As the arising oscillatory coefficients become highly non-trivial functions of $\tau_Q$, we show the exact calculations only for the average,
\begin{eqnarray}
    &&\kappa^{(\log)}_1=\sum_{k>0}\log\left|2p_k-\cos\Theta_k\right|-\log\left|\cos\Theta_k\right|\\
    &&+\log\left|2\sqrt{p_k(1-p_k)}\sin(2\tau_Q)\right|+\log\left|\sin\Theta_k+2\sqrt{p_k(1-p_k)}\cos(2\tau_Q)\right|-\log\left|\sin\Theta_k\right|\nonumber\\
    &&\approx\frac{L}{2\pi}\int_0^\pi\mathrm dk\log\left|\frac{2p_k}{\cos\Theta_k}-1\right|+\log\left|1+\frac{2\sqrt{p_k(1-p_k)}}{\sin\Theta_k}\cos(2\tau_Q)\right|+\frac{1}{2}\log|1-p_k|+\frac{1}{2}\sum_{k>0}\log(2\sin(2\tau_Q)p_k),
\end{eqnarray}
where the last term arises because the corresponding ground-state Pauli spectrum value is zero, leading to a divergent constant factor. For this reason and to give a better characterization of the cumulants, we focus on the well-defined converging logarithmic values, which give
\begin{eqnarray}
    \kappa^{(\log)}_1\approx \frac{L}{2\pi}\int_0^\infty\mathrm dx\left[\log\left|1-2p(x)\right|+\log\left|1+\tau^{1/2}_Q\frac{2\sqrt{p(x)(1-p(x))}}{C_\beta x^{\beta-1}}\cos(2\tau_Q)\right|+\frac{1}{2}\log|1-p(x)|\right]\tau^{-\frac{1}{2(\beta-1)}}_Q,
\end{eqnarray}
Note that even though an additional $\tau^{1/2}_Q$ factor and a dangerous denominator of $x^{\beta-1}$ appears the former only provides a logarithmic correction compared to the overall $\tau^{-\frac{1}{2(\beta-1)}}_Q$ scaling, while the latter does not induce any strong deviations as it is perfectly balanced by the $\sqrt{1-p(x)}$ term for small values of $x$. For the TFIM and LRKMs, this is confirmed by the small momentum $k^{2(\beta-1)}\tau_Q\ll1$ expansion of the LZ transition probabilities, $\sqrt{1-p(x)}/k^{\beta-1}\propto\tau^{1/2}_Q(1+O(k^{\beta-1}\tau^{1/2}_Q))$. 

\section{Numerical technique to construct the  statistics of the logarithmic Pauli spectrum}

To efficiently construct the histograms of the logarithmic Pauli spectrum for large system sizes $L=200$ we build the histogram counts iteratively in the following way. With a sufficiently small bin size and large enough total sample width, we store the first four histogram counts for the lowest momentum mode $k_0=\pi/L$, for the logarithms of the matrix elements,
\begin{eqnarray}\label{eq: Pauli_k_spec}
    \Sigma^{(1)}_{k}=\langle\psi_k\lvert \sigma^0_k\sigma^0_{-k}\rvert\psi_k\rangle&=&\langle\psi_k\lvert \sigma^z_k\sigma^z_{-k}\rvert\psi_k\rangle=1,\\
    \Sigma^{(2)}_{k}=\langle\psi_k\lvert \sigma^z_k\sigma^0_{-k}\rvert\psi_k\rangle&=&\langle\psi_k\lvert \sigma^0_k\sigma^z_{-k}\rvert\psi_k\rangle=\lvert u_k\rvert^2-\rvert v_k\lvert^2\approx2p_k-\cos \Theta_k,\nonumber\\
    \Sigma^{(3)}_{k}=\langle\psi_k\lvert \sigma^x_k\sigma^y_{-k}\rvert\psi_k\rangle&=&\langle\psi_k\lvert \sigma^y_k\sigma^x_{-k}\rvert\psi_k\rangle=-2\,\mathrm{Im}\{u_k v^*_k\}\approx-2\sqrt{p_k(1-p_k)}\sin\Phi_k,\nonumber\\
    \Sigma^{(4)}_{k}=\langle\psi_k\lvert \sigma^x_k\sigma^x_{-k}\rvert\psi_k\rangle&=&2\mathrm{Re}\{u_kv^*_k\}=-\langle\psi_k\lvert \sigma^y_k\sigma^y_{-k}\rvert\psi_k\rangle\approx\sin\Theta_k+2\sqrt{p_k(1-p_k)}\cos\Phi_k,\nonumber
\end{eqnarray}
denoted by $y^{(j)}_0\equiv \log \lvert\Sigma^{(j)}_{k_0}\rvert,j=1,\dots,4$ in the discrete distribution function of 
\begin{eqnarray}
    P_0(Y)=\sum_{j=1}^4\delta(Y-y^{(j)}_0).
\end{eqnarray}

For the second TLS, $k_1=3\pi/L$, the new values $y^{(j)}_1=\log\Sigma^{(j)}_{k_1}$ ($j=1,\dots,4$) are added to the previous ones by which the new histogram counts become
\begin{eqnarray}
    P(Y_1)=\sum_{j,j'=1}^4\delta(Y_1-y^{(j)}_1-y^{(j)}_0).
\end{eqnarray}
This step only requires a step of $4\times N_\mathrm{bin}$ ($N_\mathrm{bin}$ denotes the total number of counting bins), where the $i$ histogram count is updated by $W_{i,0}\rightarrow W_{i,0}+w_{i,1}$ repeated for all the four values. Generally, $W_{i,n}$ is the number of counts in the $i$th bin, after $n$ steps, while $w_{i,n}$ is the number of counts only from the $n$-th TLS.
The iteration simply follows then for any $n$-th step as having already constructed the histogram $P(Y_{n-1})=\sum_{j_0,\dots j_{n-1}}\delta(Y_{n-1}-y^{(j_0)}_{k_0}-\dots y^{(j_{n-1})}_{k_{n-1}})$ with the bin weights again updated as $W_{i,n-1}\rightarrow W_{i,n-1}+w_{i,n}$. Thus, in total, the method requires only $4L\times N_\mathrm{bins}$ numerical steps, instead of the naive estimate $4^L$ for storing all possible values of the Pauli spectrum. The distribution of the original Pauli spectrum values is then obtained by a change of variables, $X=e^Y$.

\section{Exact results in the TFIM for the SRE}
In the TFIM, executing leading order analysis further validates our universal prediction. In particular, the ground and excited states amplitudes read ~\cite {Moessner2022,Cincio_2007}
\begin{eqnarray}
    &&\lvert v_k\rvert^2= \sin^2\frac{k}{2}+e^{-2\pi k^2\tau_Q}\equiv\sin^2\frac{k}{2}+ p_k,\\
    &&\lvert u_k\rvert^2= \cos^2\frac{k}{2}-e^{-2\pi k^2\tau_Q}\equiv \cos^2\frac{k}{2}-p_k,\nonumber\\
    &&u_kv^*_k=\frac{1}{2}\sin k+\sqrt{p_k(1-p_k)}e^{i\Phi_k}\equiv\frac{1}{2}\sin k+e^{-\pi k^2\tau_Q}\sqrt{1-e^{-2\pi k^2\tau_Q}}e^{i\Phi_k},\nonumber\\
    &&\Phi_k=\frac{\pi}{4}+2\tau_Q+k^2\tau_Q\left[\log (4\tau_Q)+\gamma_E-2\right]\approx2\tau_Q,\nonumber
\end{eqnarray} 
which match perfectly with the general formula in Eq.~\eqref{eq: Pauli_k_spec} with $p_k=e^{-2\pi k^2\tau_Q}$ and $\Theta_k=k$. Thus,  
    \begin{eqnarray}
    &&\langle\psi_k\lvert \sigma^0_k\sigma^0_{-k}\rvert\psi_k\rangle=\langle\psi_k\lvert \sigma^z_k\sigma^z_{-k}\rvert\psi_k\rangle=1,\\
    &&\langle\psi_k\lvert \sigma^z_k\sigma^0_{-k}\rvert\psi_k\rangle=\langle\psi_k\lvert \sigma^0_k\sigma^z_{-k}\rvert\psi_k\rangle=\lvert u_k\rvert^2-\rvert v_k\lvert^2=2e^{-2\pi k^2\tau_Q}-\cos k,\\
    &&\langle\psi_k\lvert \sigma^x_k\sigma^y_{-k}\rvert\psi_k\rangle=\langle\psi_k\lvert \sigma^y_k\sigma^x_{-k}\rvert\psi_k\rangle=-2\,\mathrm{Im}\{u_k v^*_k\}=-2\sqrt{e^{-2\pi k^2\tau_Q}(1-e^{-2\pi k^2\tau_Q})}\sin(2\tau_Q),\\
    &&\langle\psi_k\lvert \sigma^x_k\sigma^x_{-k}\rvert\psi_k\rangle=2\mathrm{Re}\{u_kv^*_k\}=-\langle\psi_k\lvert \sigma^y_k\sigma^y_{-k}\rvert\psi_k\rangle=\sin  k+2\sqrt{e^{-2\pi k^2\tau_Q}(1-e^{-2\pi k^2\tau_Q})}\cos(2\tau_Q).
\end{eqnarray}
From here, the general form of the integral function, Eq.~\eqref{eq: I_alpha} reads
\begin{eqnarray}
    I_{\{p(x)\}}(\alpha,\tau_Q)=\int_0^\infty\mathrm dx\,\log_2\left[
    \frac{
1+\left(1-2e^{-2\pi x^2}\right)^{2\alpha}
+
\left(4e^{-2\pi x^2}(1-e^{-2\pi x^2})\right)^\alpha
\left(\sin^{2\alpha}(2\tau_Q)+\cos^{2\alpha}(2\tau_Q)\right)
}{2}
    \right].\nonumber\\
\end{eqnarray}
Further numerical verifications of the TFIM and LRKMs across various scaling regimes are provided in the following section.

\section{Further numerical demonstrations in the TFIM and in the LRKMs}
\begin{figure}[h!]
\includegraphics[width=.49\linewidth]{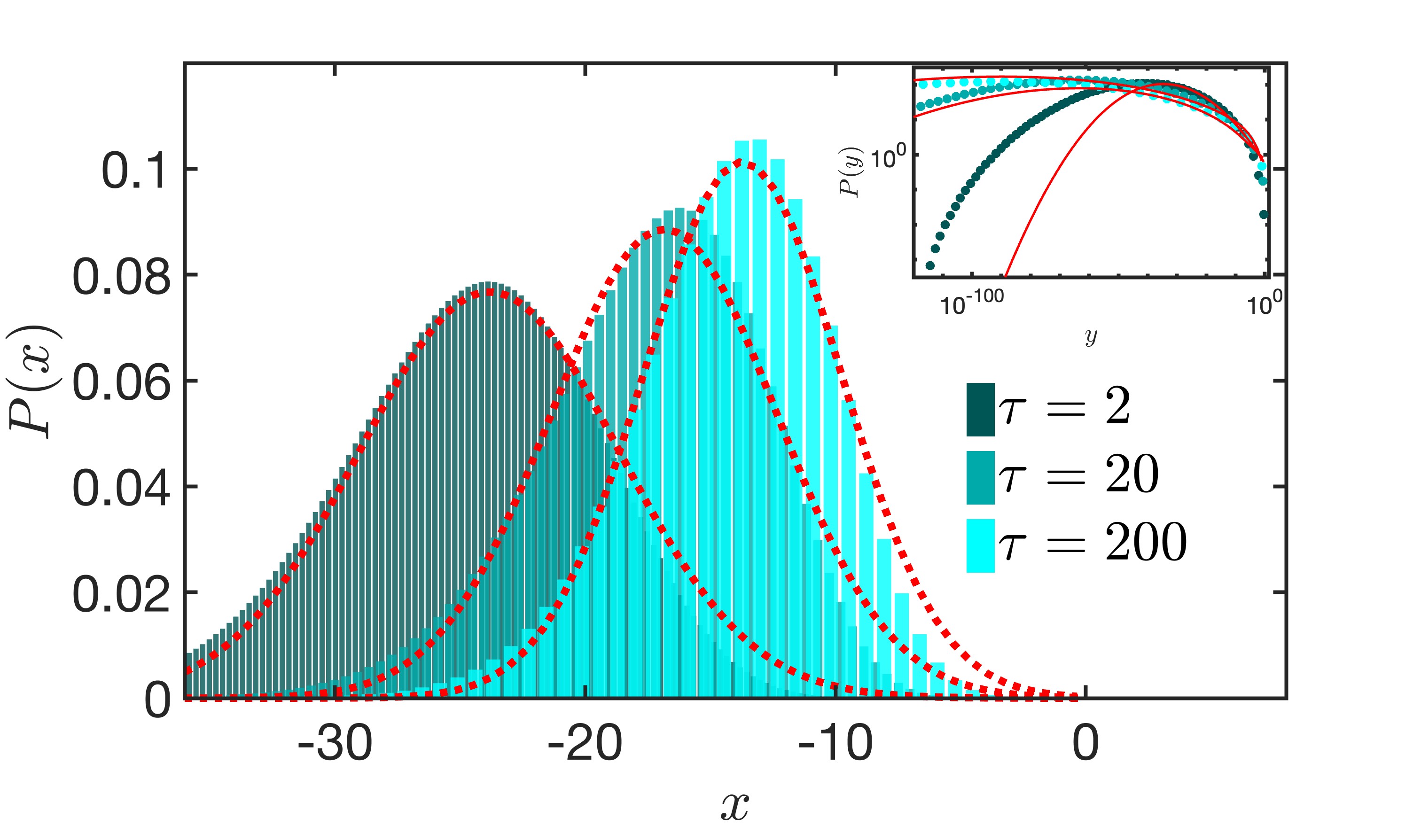}\\
    \caption{Distribution of the logarithmic Pauli spectrum values in the TFIM following the predicted Gaussian character ($L=200$). The histograms approach the final ground-state values as the width decreases with increasing $\tau_Q$. The inset shows the original statistics of the Pauli spectrum, following the predicted lognormal distribution.}
    \label{fig:TFIM_Stat}
\end{figure}

\begin{figure}[h!]
\includegraphics[width=.49\linewidth]{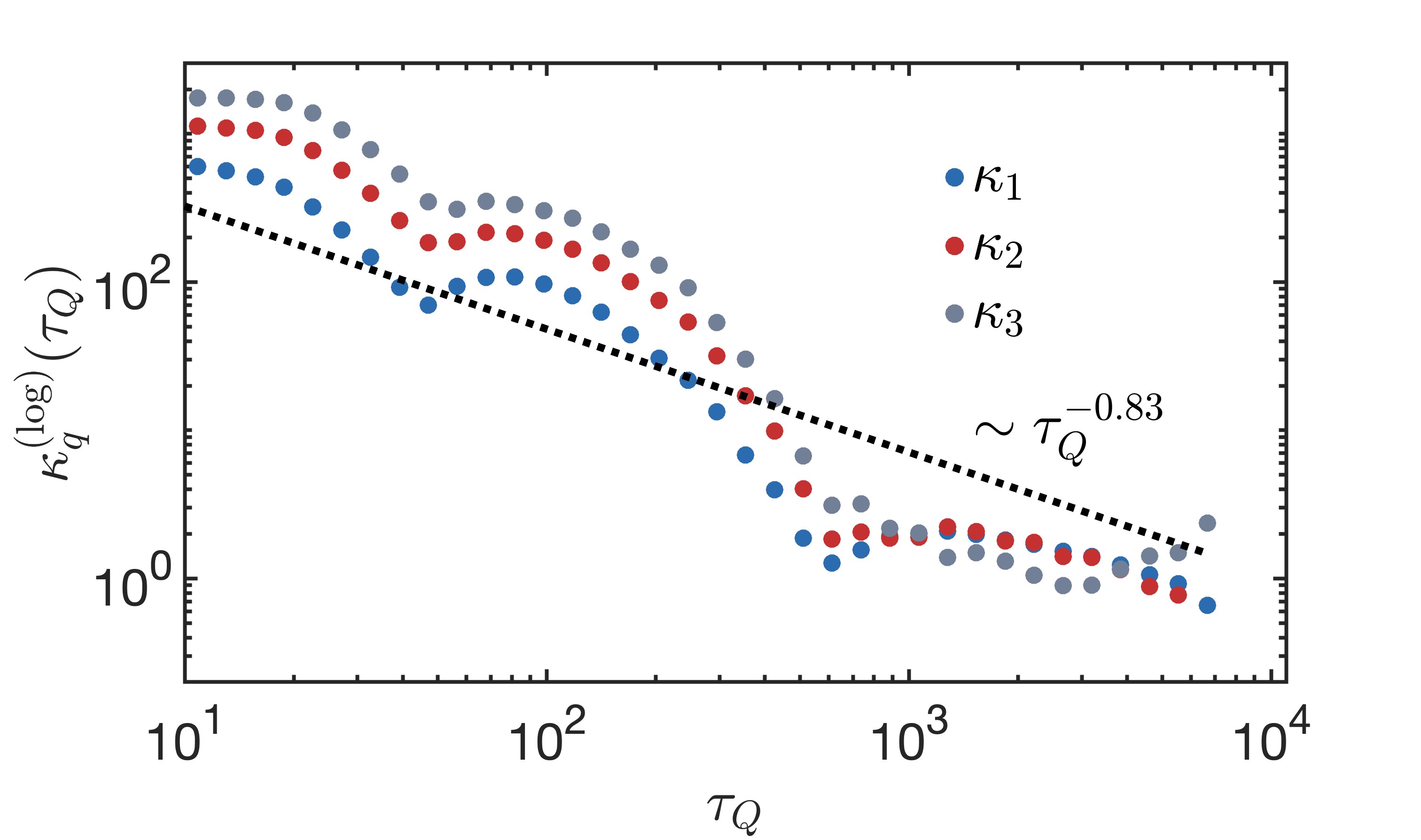}\\
    \caption{Cumulants of the logarithmic Pauli spectrum for $\gamma=1.4,\,\beta=1.6$ in the LRKM following the predicted dynamical power-law scaling up to slight oscillations.}
    \label{fig:LRKM_cum_a1_4b1_6}
\end{figure}

In this section, we first show that the histogram counts of the logarithmic Pauli spectrum and the Pauli spectrum itself follow the predicted Gaussian and lognormal distributions, respectively, as shown in Fig.~\ref{fig:TFIM_Stat}.
Next, we show that the cumulant scaling of the logarithmic Pauli spectrum for the LRKMs in the dynamical scaling regime, $\gamma=1.4,\,\beta=1.6$, follows the predicted power-law $\kappa^{(\log)}_q\propto\tau^{-\frac{1}{2(\beta-1)}}_Q\propto\tau^{-0.83}_Q$, as shown in Fig.~\ref{fig:LRKM_cum_a1_4b1_6}.

\begin{figure}[h!]
\includegraphics[width=.32\linewidth,trim={2.5cm 0 7cm 0},clip]{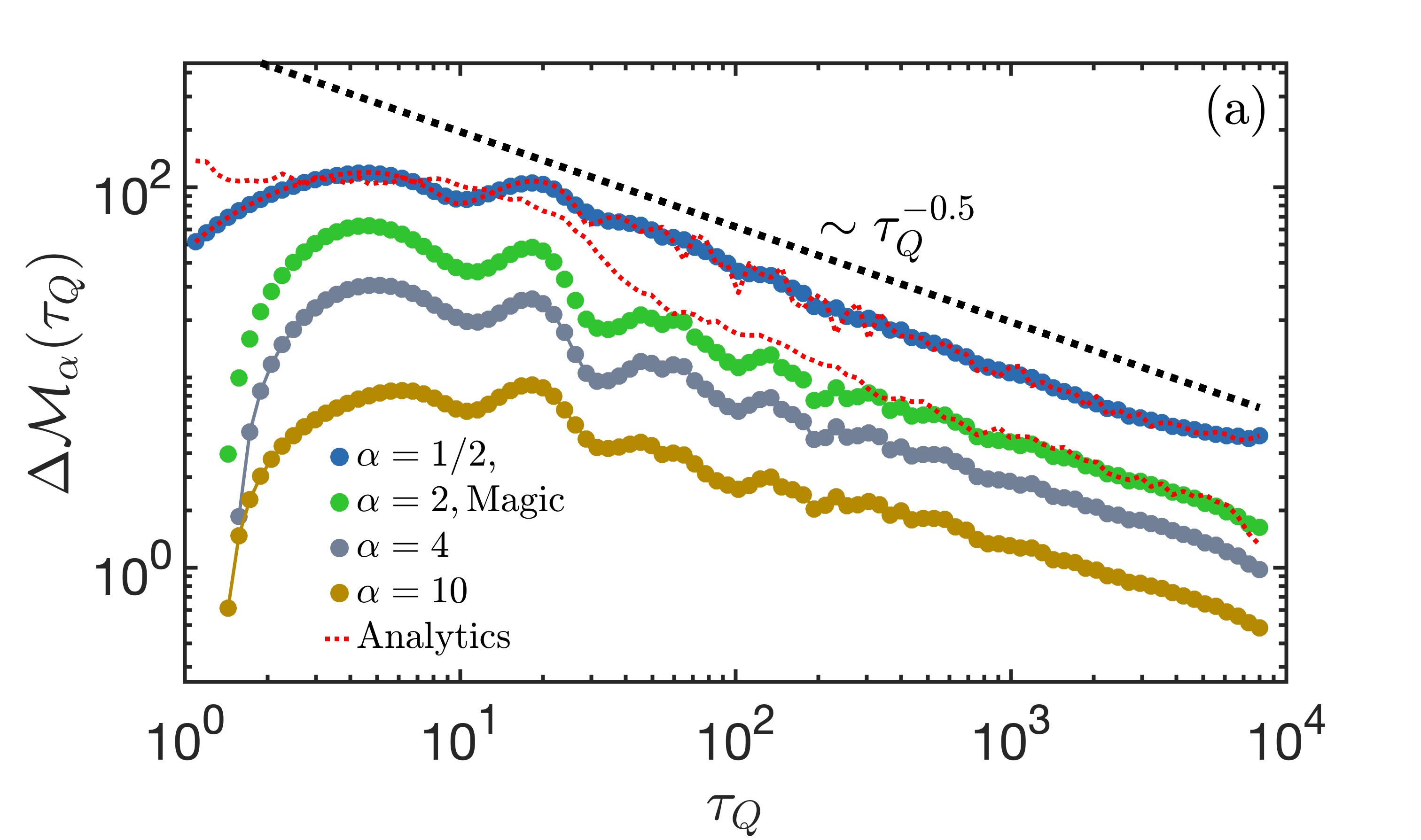}
\includegraphics[width=.32\linewidth,trim={2.5cm 0 7cm 0},clip]{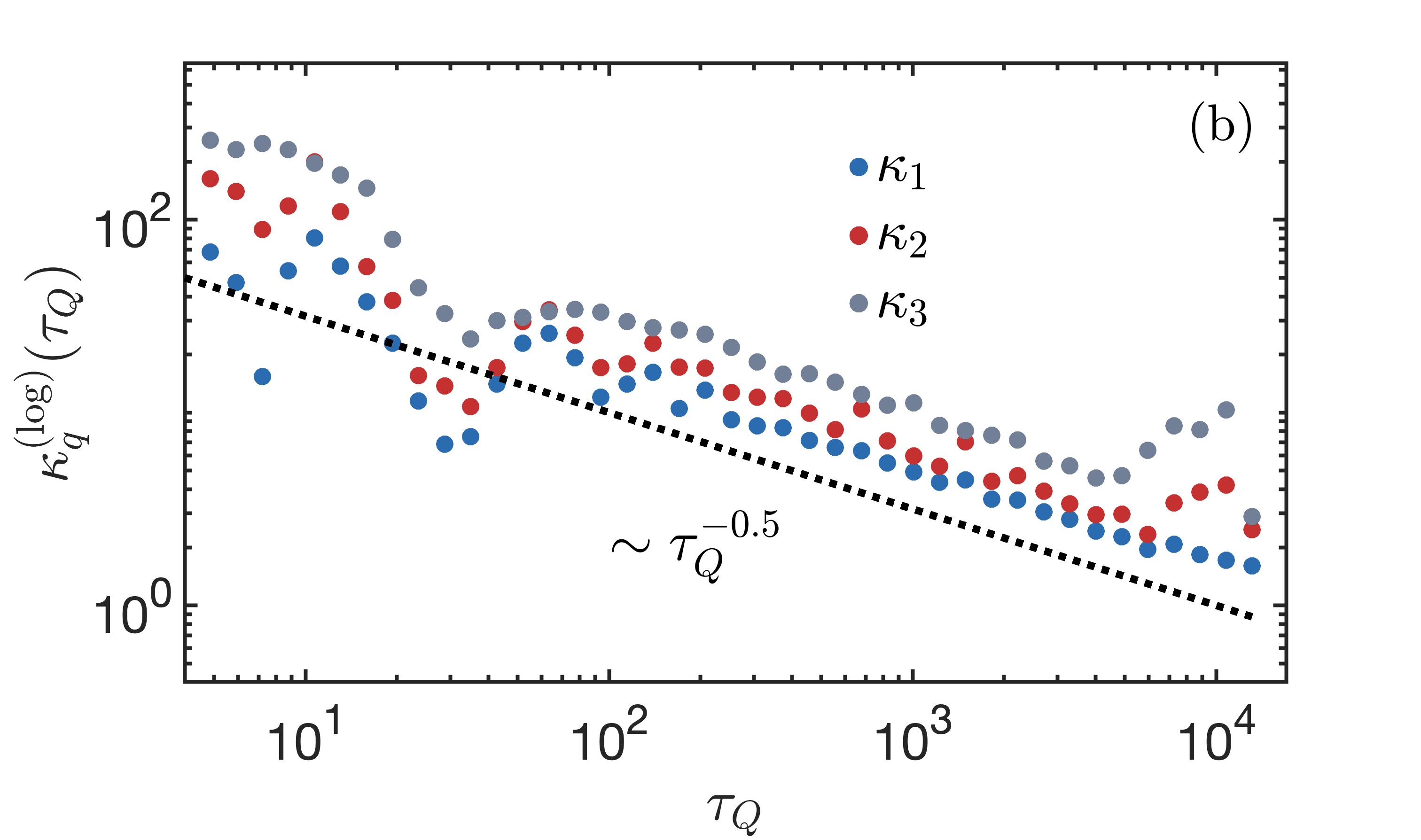}
\includegraphics[width=.32\linewidth,trim={1.cm 0 10cm 0},clip]{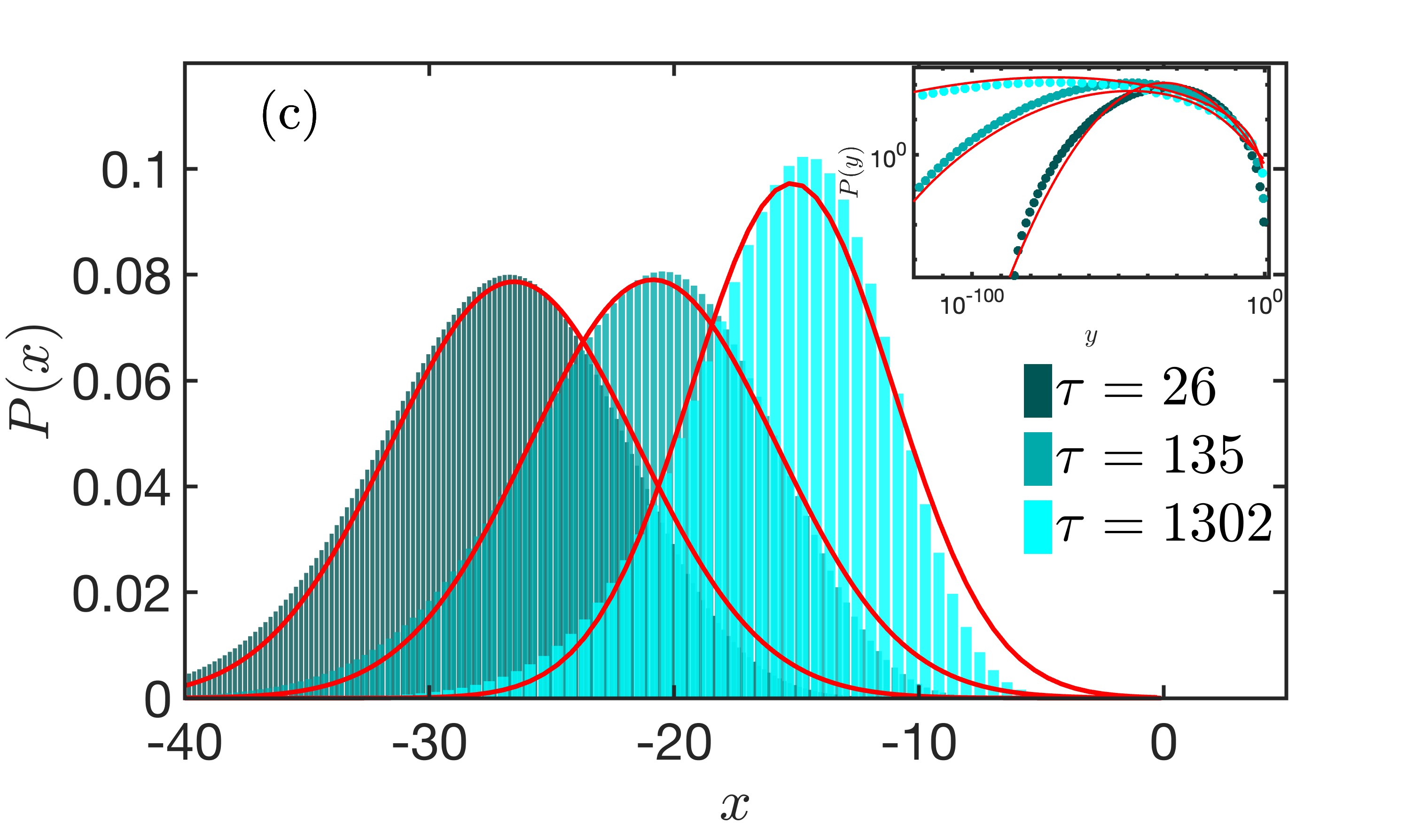}
    \caption{(a) Stabilizer R\'enyi entropy for quantum magic, $\alpha=1/2$, and higher order SREs in the LRKM ($\alpha=4,10$), following the same universal power-laws. (b) Cumulants of the logarithmic Pauli spectrum following the predicted $\tau^{-1/2}_Q$ scaling for $\gamma,\,\beta=5$.  (c) Histogram counts of the logarithmic Pauli spectrum values following the predicted Gaussian character, with the inset showing the lognormal distribution fits for the Pauli spectrum statistics ($L=200$).}
    \label{fig:LRKM_a5_b5}
\end{figure}

We also cover all other scaling regimes of the LRKMs and show the corresponding scalings of the cumulants, as well as the Gaussian and lognormal behavior of the Pauli spectrum statistics. For $\gamma,\,\beta=5$, similar results are obtained as in the TFIM: the first three cumulants decay as $\tau^{-1/2}_Q$ and the corresponding histograms follow precisely the Gaussian and lognormal distributions as shown in Fig.~\ref{fig:LRKM_a5_b5}. 

\begin{figure}[h!]
\includegraphics[width=.32\linewidth,trim={2.5cm 0 7cm 0},clip]{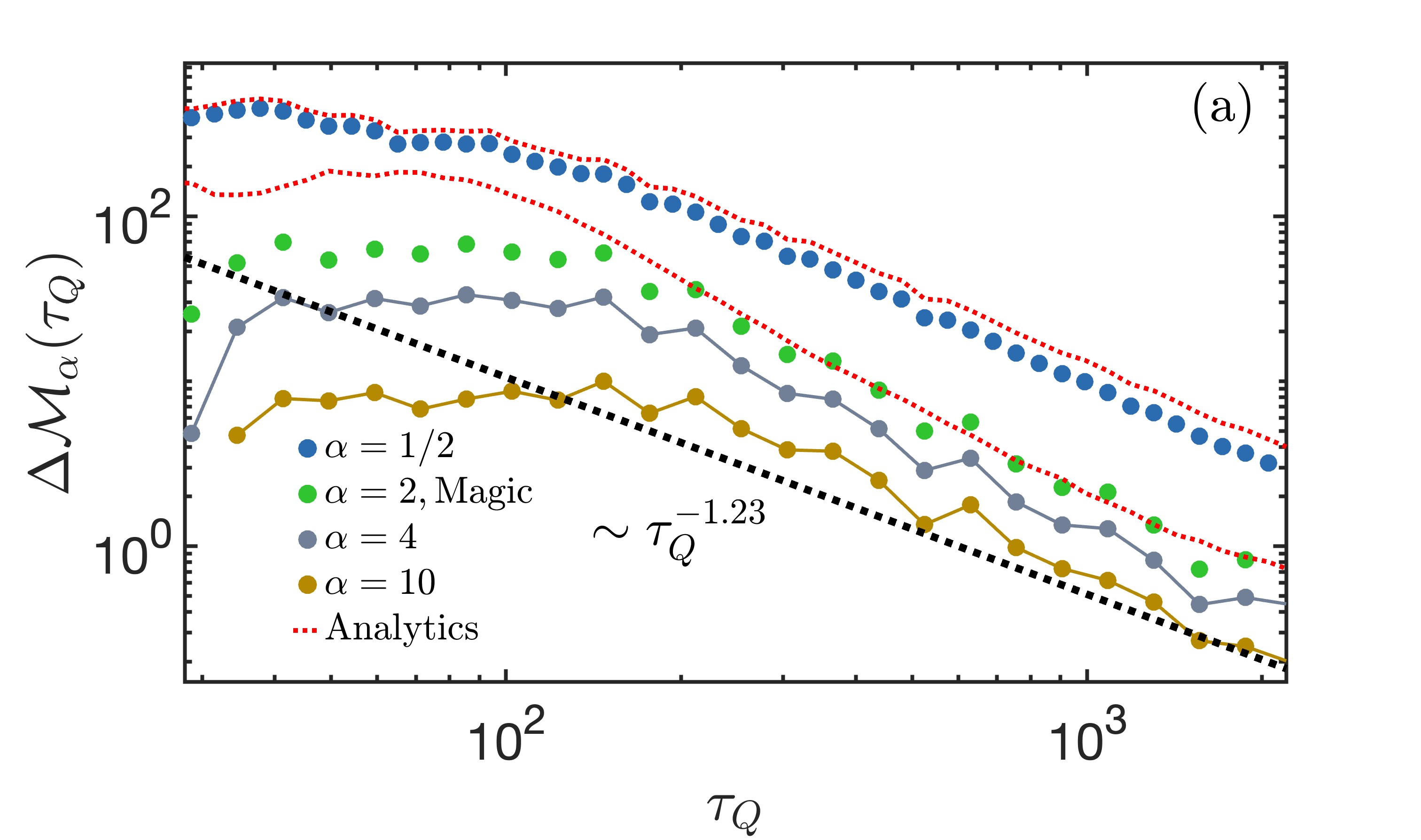}
\includegraphics[width=.32\linewidth,trim={2.5cm 0 7cm 0},clip]{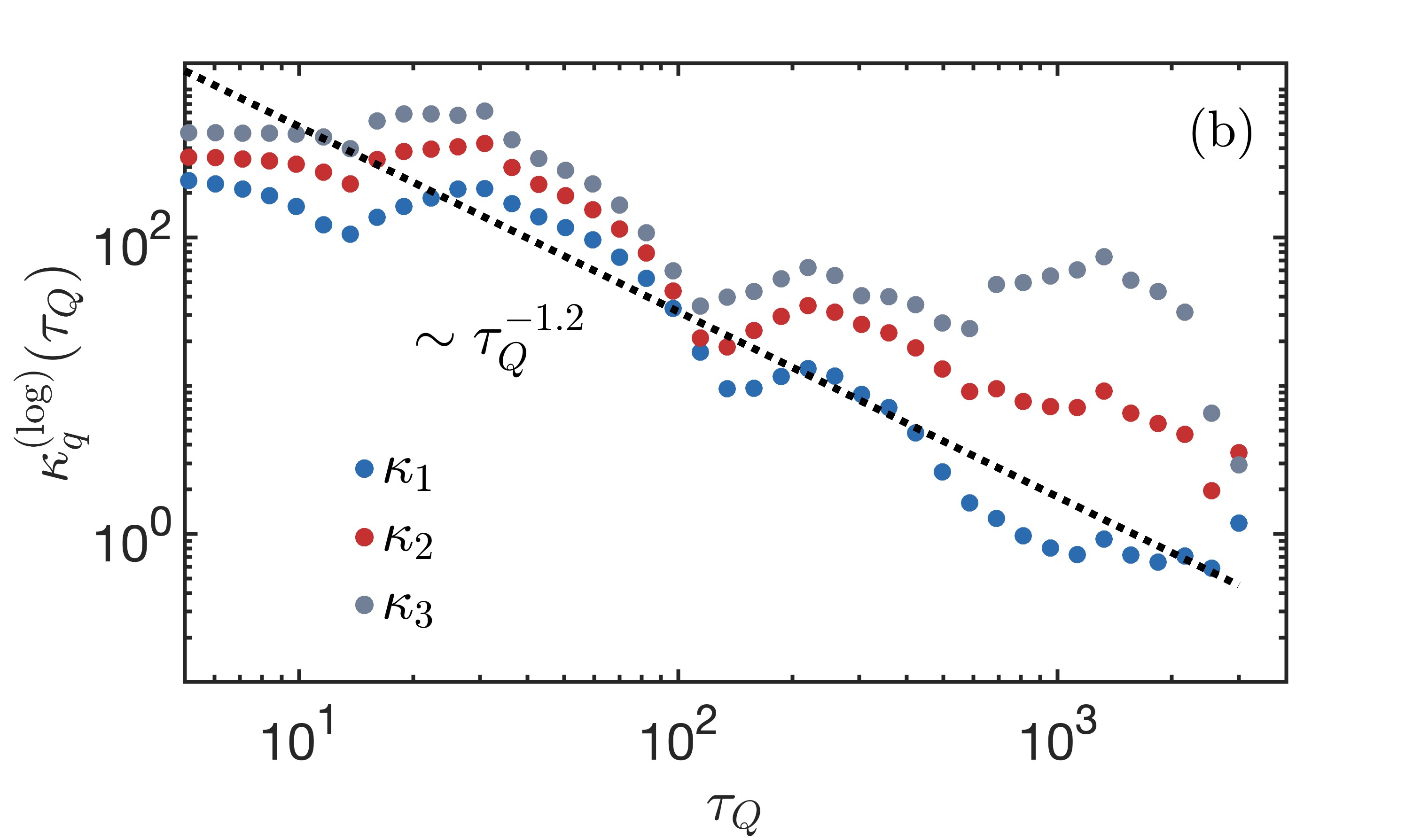}
\includegraphics[width=.32\linewidth,trim={1.cm 0 10cm 0},clip]{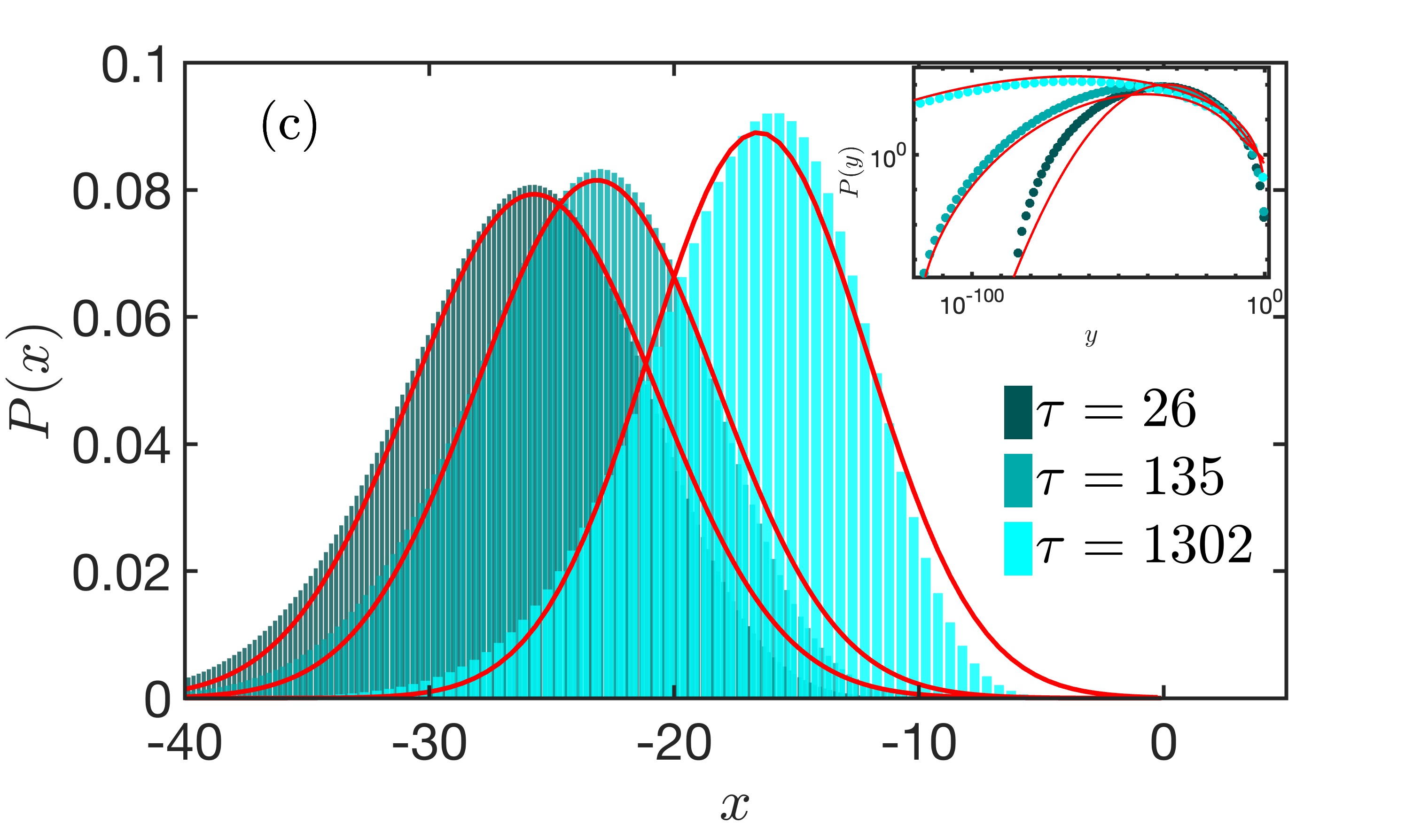}
    \caption{(a) Stabilizer R\'enyi entropy for quantum magic, $\alpha=1/2$ and higher order SREs in the LRKM  ($\alpha=4,10$), following the same universal power-laws as the cumulants of the logarithmic Pauli spectrum in panel (b)  $\tau^{-1.2}_Q$ for $\gamma=2.5,\,\beta=1.4$. (c) The statistics of the Pauli spectrum and its logarithmic version are in good agreement with the log-normal and Gaussian limiting forms.}
    \label{fig:LRKM_a2_5_b1_4}
\end{figure}

The numerical findings for the short-range regime, where the KZ scaling still holds, $\gamma=2.5,\beta=1.4$, exhibit analogous features. As shown in Fig.~\ref{fig:LRKM_a2_5_b1_4}, both the SRE and the cumulants of the logarithmic Pauli spectrum follow the power-law $\kappa^{(\log)}_q\propto\tau^{-\frac{1}{2(\beta-1)}}_Q\propto\tau^{-1.25}_Q$, with the same lognormal and Gaussian statistics for the histogram counts of the Pauli spectrum and its logarithmic version.

\begin{figure}[t!]
\includegraphics[width=.32\linewidth,trim={2.5cm 0 7cm 0},clip]{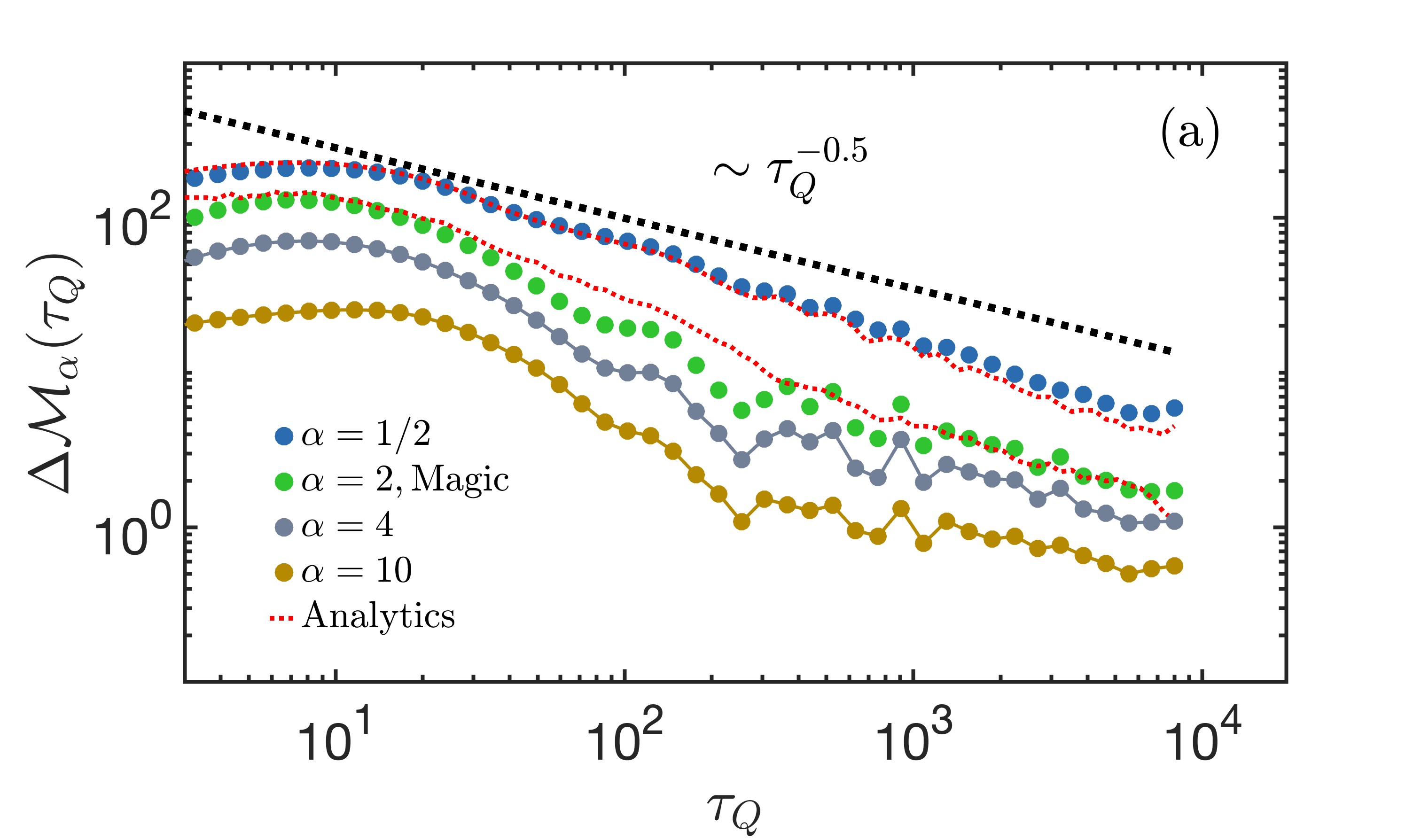}
\includegraphics[width=.32\linewidth,trim={2.5cm 0 7cm 0},clip]{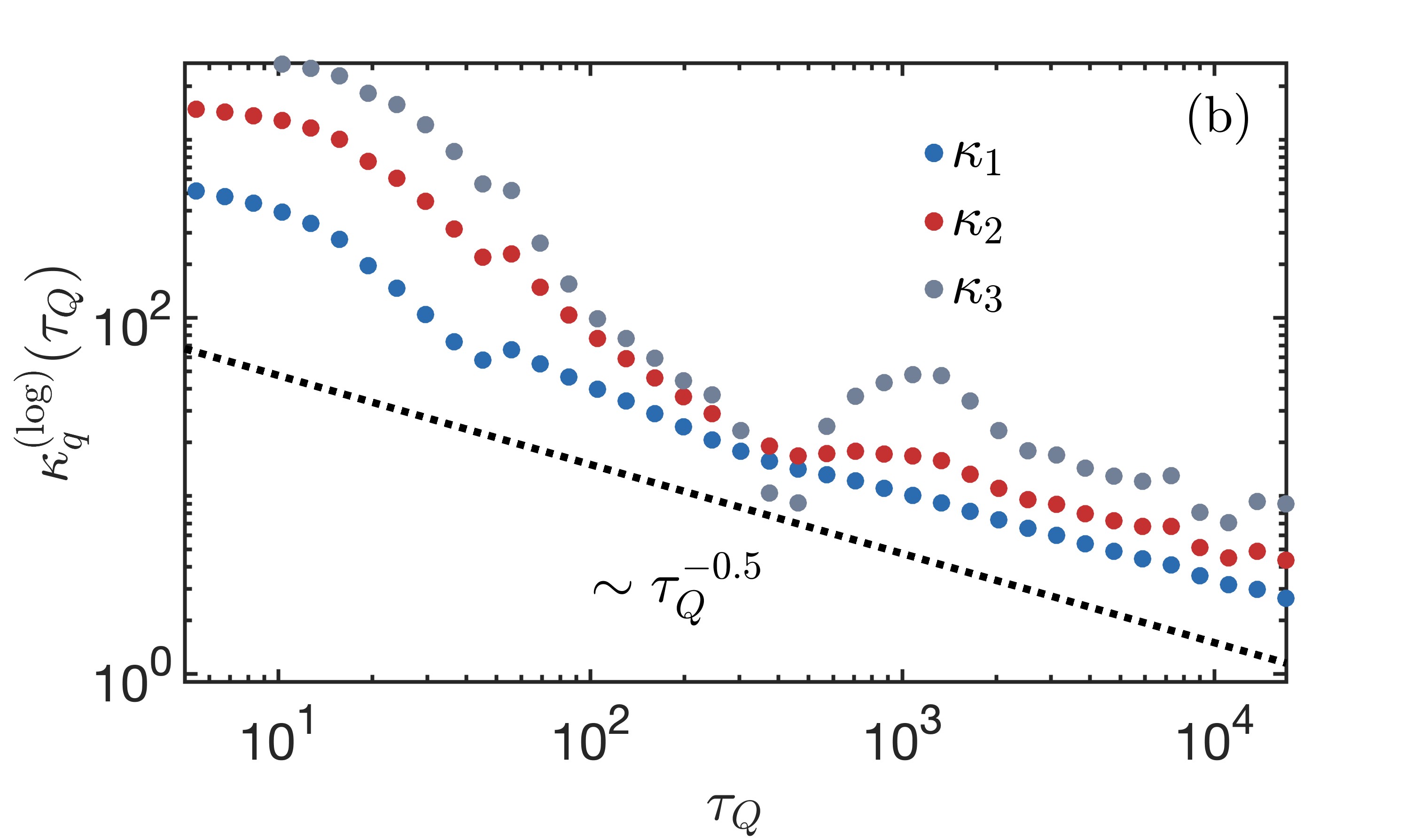}
\includegraphics[width=.32\linewidth,trim={1.cm 0 10cm 0},clip]{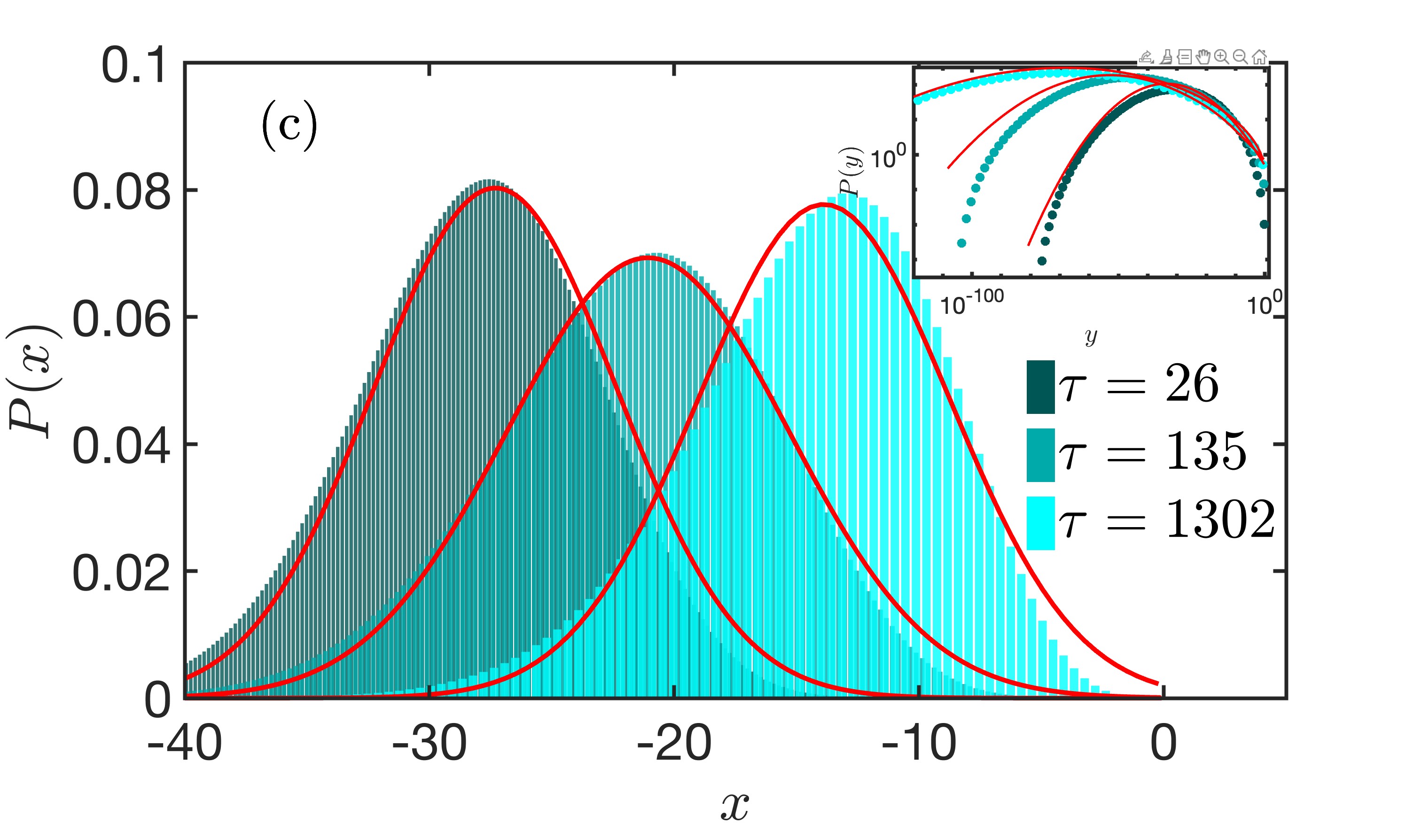}
    \caption{(a) Stabilizer R\'enyi entropy for quantum magic, $\alpha=1/2$ and higher order SREs in the LRKM, ($\alpha=4,10$), and (b) cumulants of the logarithmic Pauli spectrum decaying as  $\tau^{-1/2}_Q$ for $\gamma=1.2,\,\beta=5$. (c) Lognormal and Gaussian statistics for the Pauli spectrum and the corresponding logarithmic values.}
    \label{fig:LRKM_a1_2_b5}
\end{figure}

Finally, all the above observations hold for the short-range pairing, $\beta=5$ and long-range hopping, $\gamma=1.2$ cases with the universal power-law given by $\Delta\mathcal M_\alpha\propto\kappa^{(\log)}_q\propto \tau^{-1/2}_Q$ and with the same features for the statistics of the Pauli spectrum and its logarithmic values, as shown in Fig.~\ref{fig:LRKM_a1_2_b5}.

\section{Universal time-evolution of the SRE in the LRKMs}
\begin{figure}[t!]
\includegraphics[width=.49\linewidth,trim={2.5cm 0 7cm 0},clip]{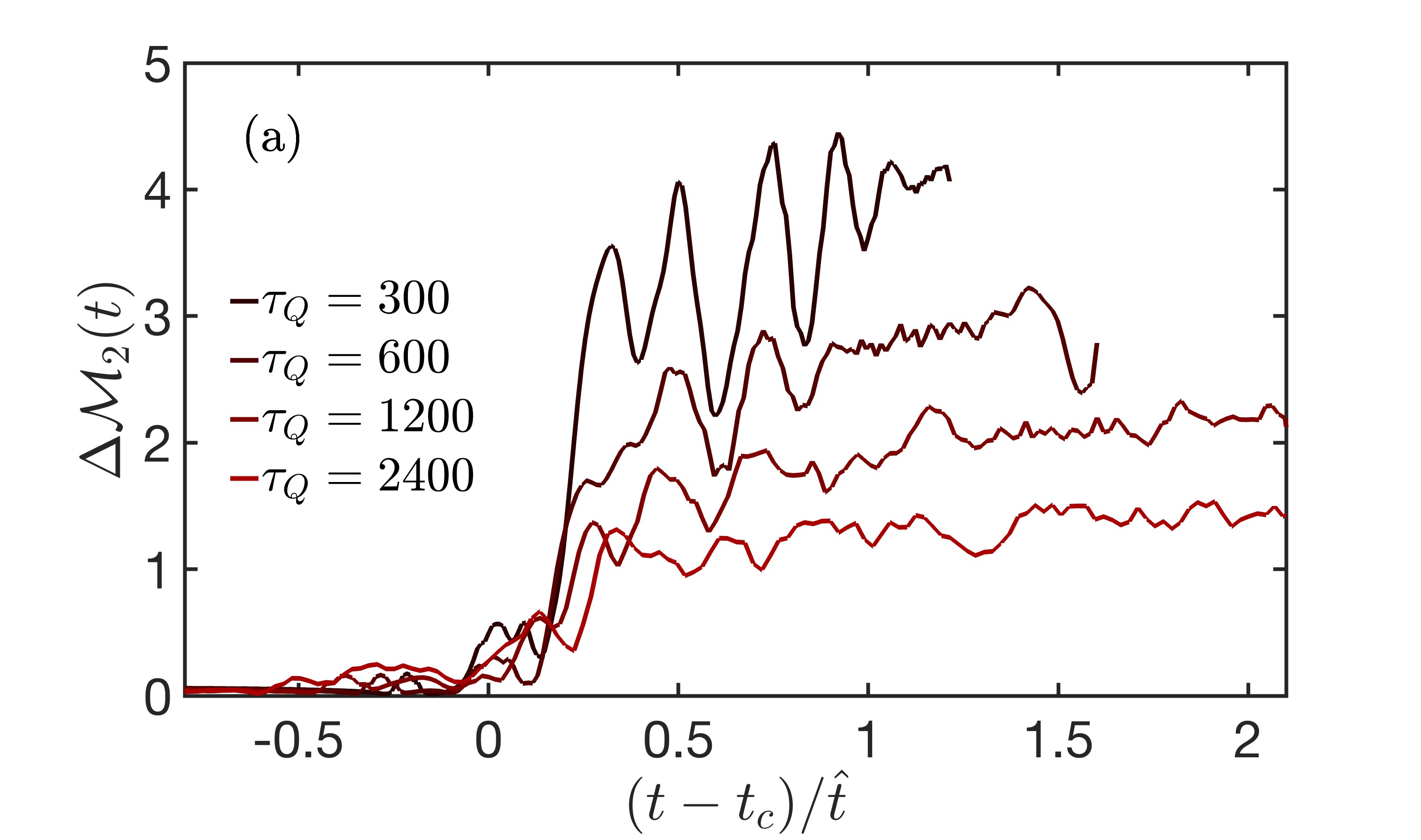}
\includegraphics[width=.49\linewidth,trim={2.5cm 0 7cm 0},clip]{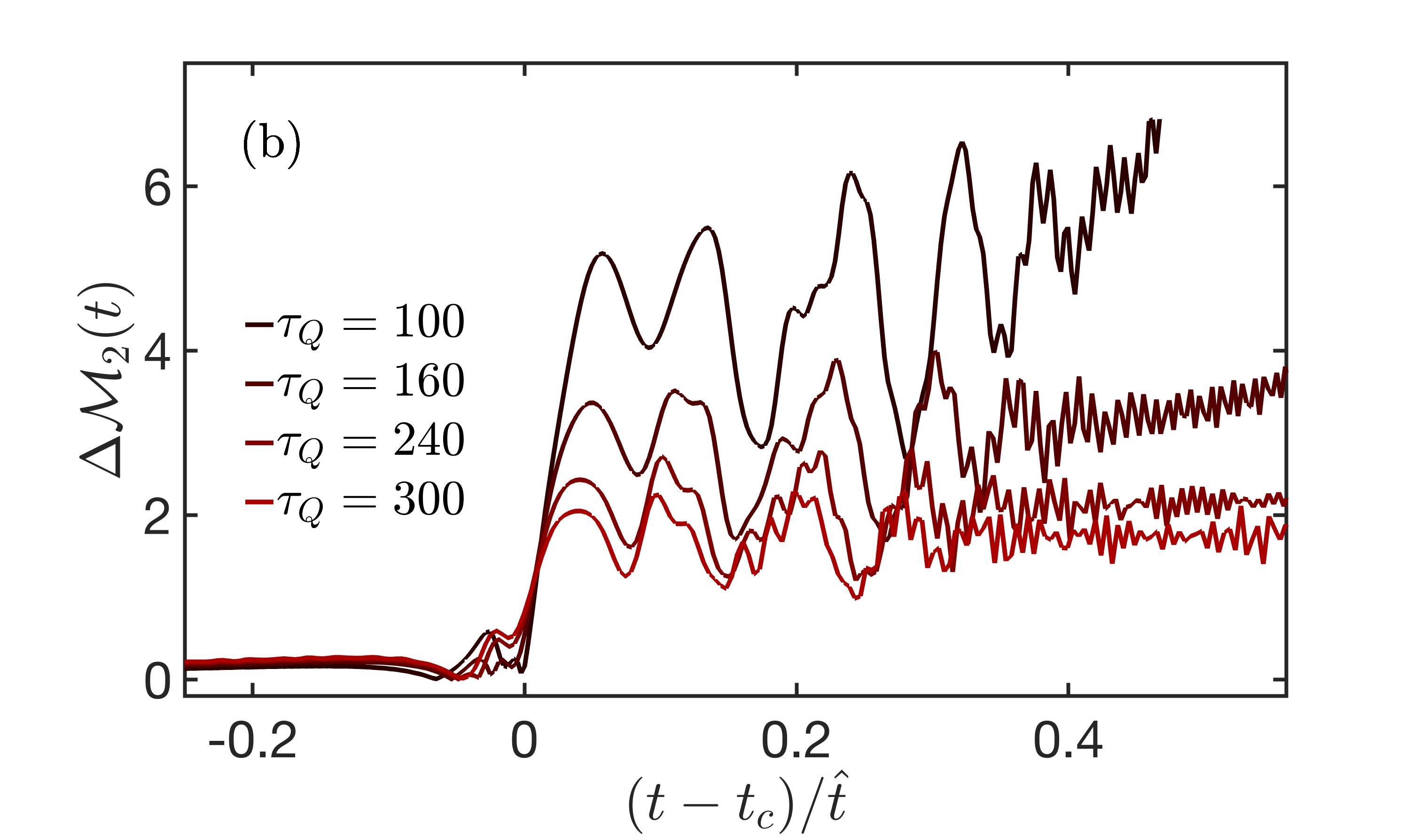}
    \caption{(a) Time evolution of the relative SRE in the LRKM in the dynamical scaling regime for short-range pairing, $\beta=5$, and long-range hopping, $\gamma=1.2$. Around $t_c$, the curves exhibit a sudden increase and collapse onto a universal scaling form upon rescaling time by $\hat t$, followed by a nonuniversal oscillatory regime.
    (b) Similar behavior with both long-range pairings and long-range hoppings, $\gamma=1.4,\,\beta=1.6$.}
    \label{fig:LRKM_TimeEvol}
\end{figure}

In this section, we further substantiate the universal near-critical dynamics of the relative SRE in the LRKMs. Using the generalized freeze-out time $\hat t$, Eqs.~\eqref{eq: t_hat} and \eqref{eq: t_hat_LRKM}, the time-evolution of the relative SRE for different $\tau_Q$ collapses when plotted versus $(t-t_c)/\hat t$ around the critical point. As shown in Fig.~\ref{fig:LRKM_TimeEvol}, the universal signatures persist even in the dynamical scaling regimes where the standard KZM prediction for defect production no longer applies. Away from the neighborhood of the critical point, the dynamics cross over to an intermediate, nonuniversal regime dominated by oscillations, which gradually relax toward the final values.

\section{Stabilizer R\'enyi entropies as a function of $\alpha$}
\begin{figure}[h!]
\includegraphics[width=.5\linewidth]{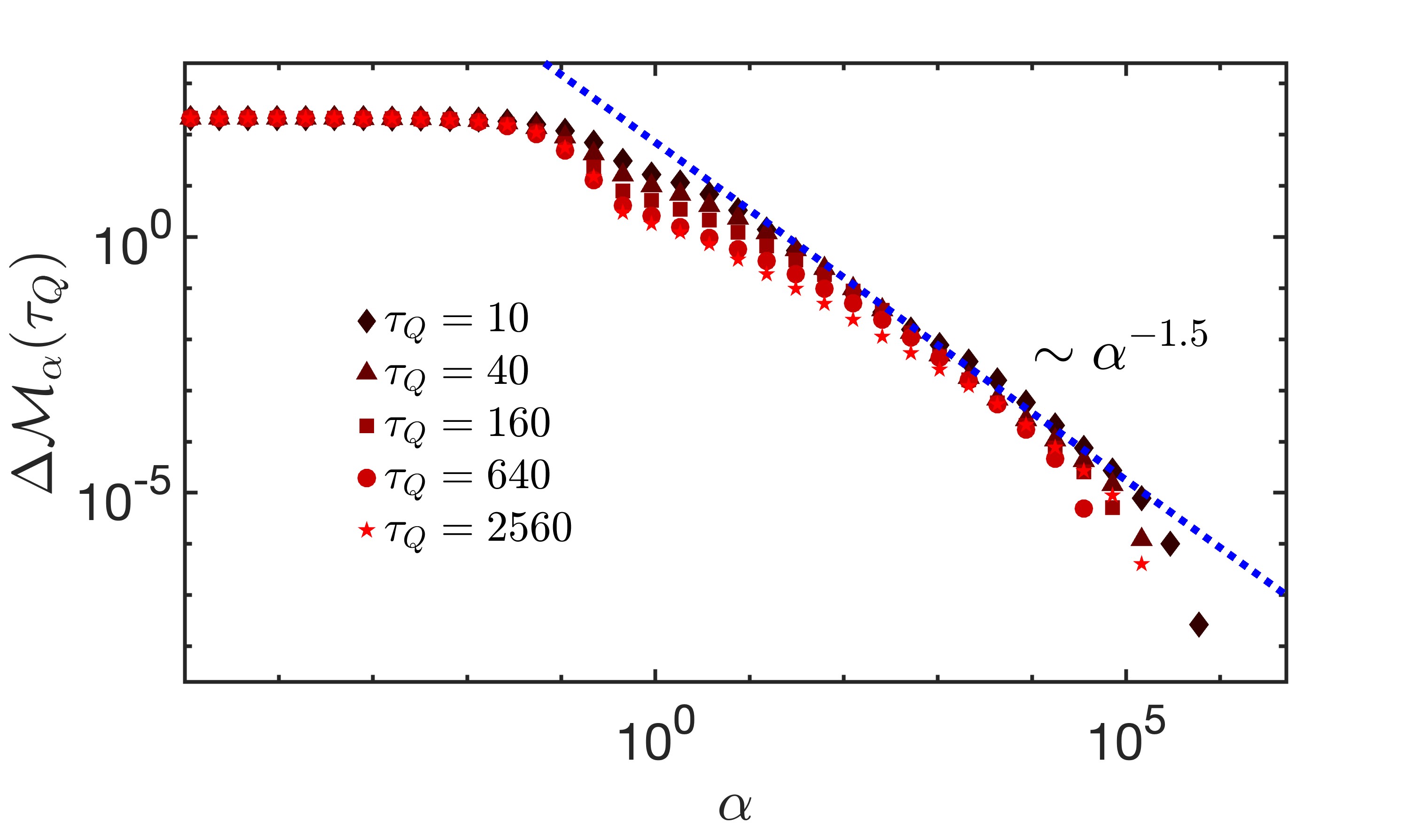}
    \caption{Stabilizer R\'enyi entropy in the TFIM as a function of $\alpha$ for fixed values of $\tau_Q$ in the slow driving regime, following precisely the predicted $\alpha^{-3/2}$ decay for large $\alpha$, while converging to the predicted constant for small $\alpha$.}
    \label{fig:TFIM_Malpha_alpha}
\end{figure}

\begin{figure}[t]
  \centering
  \includegraphics[width=0.53\textwidth,trim={0 5.5cm 7cm 4cm},clip]{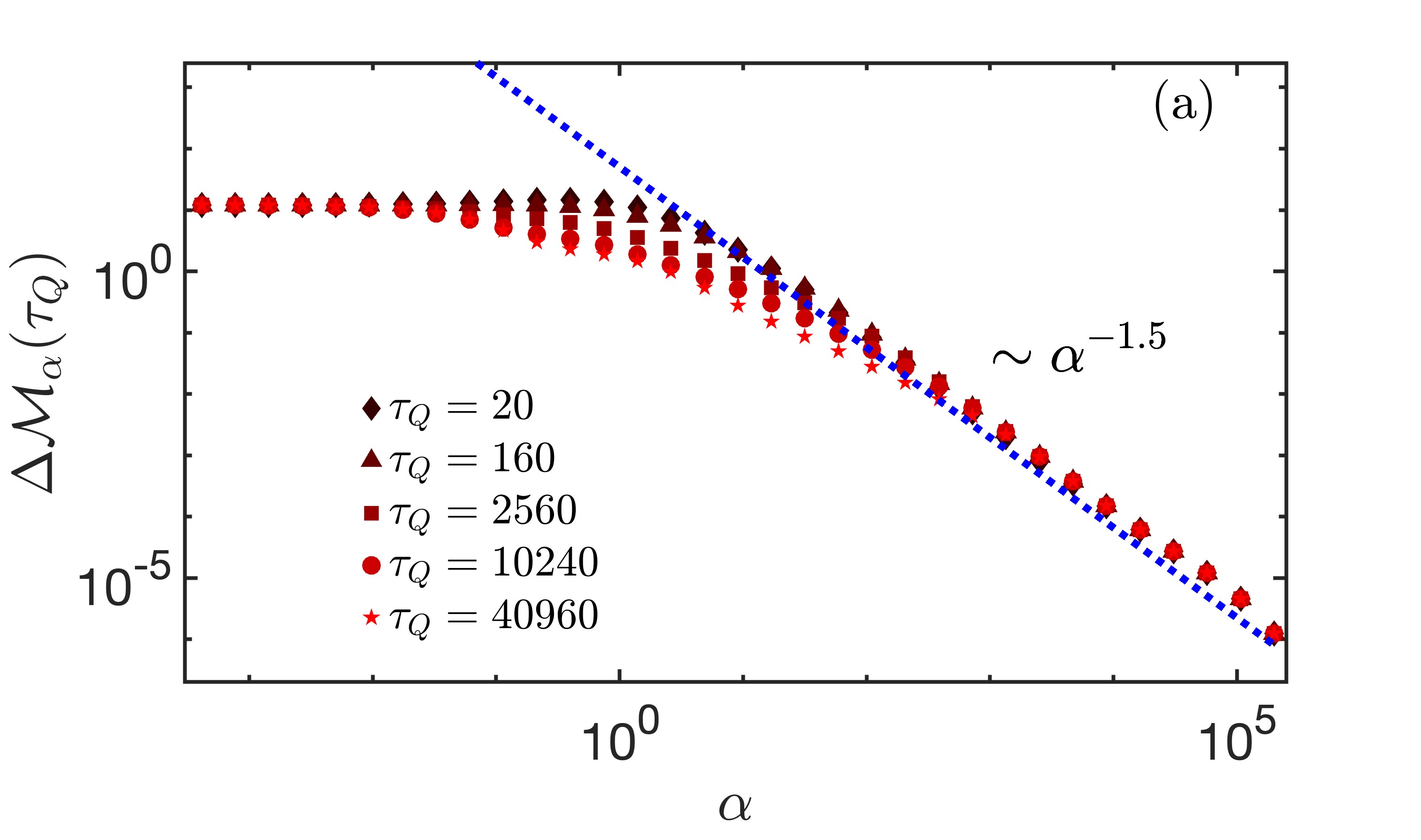}\hfill
  \includegraphics[width=0.47\textwidth,trim={13cm 5.5cm 7cm 4cm},clip]{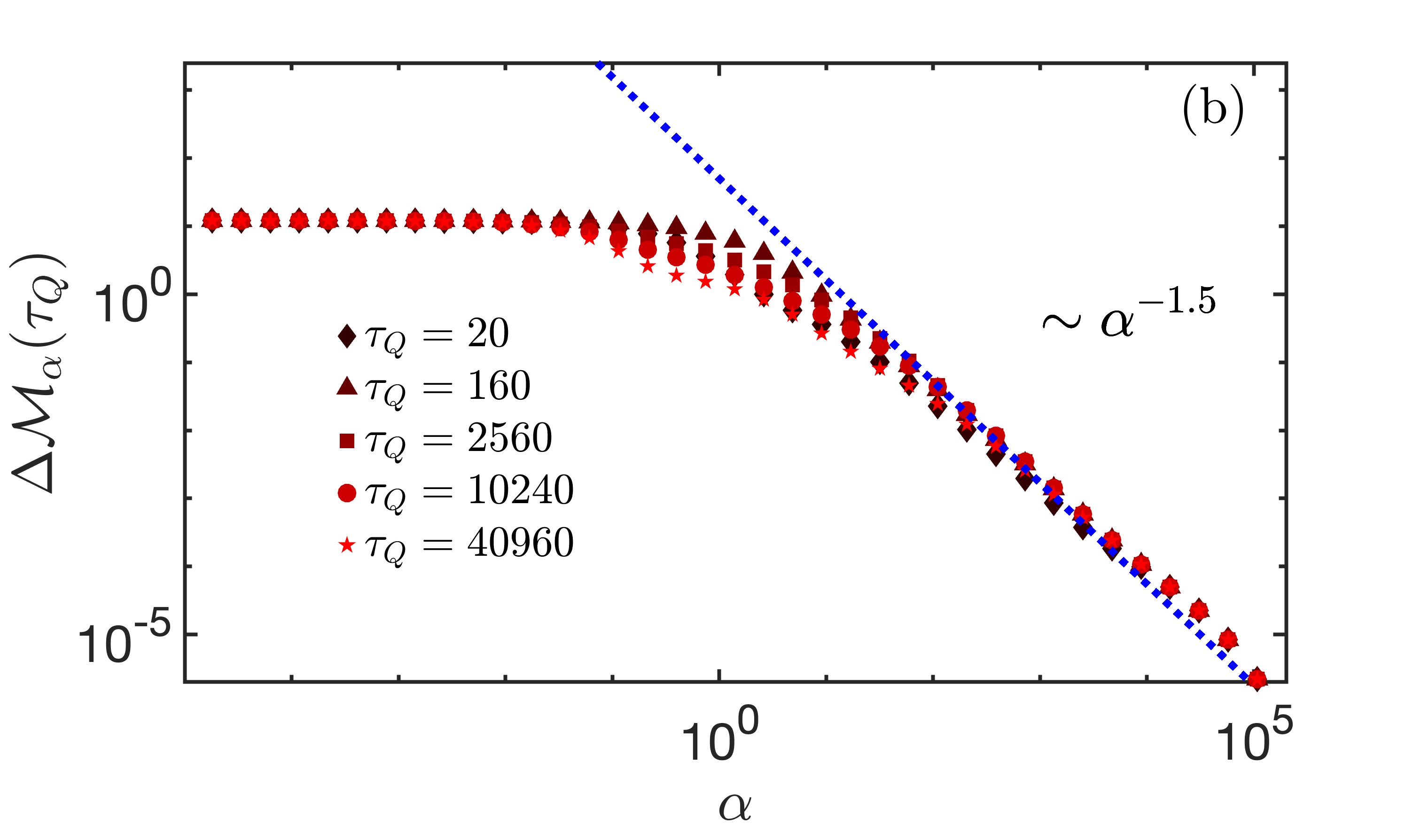}\\[1em]
  \includegraphics[width=0.53\textwidth,trim={0 0 6.5cm 4.5cm},clip]{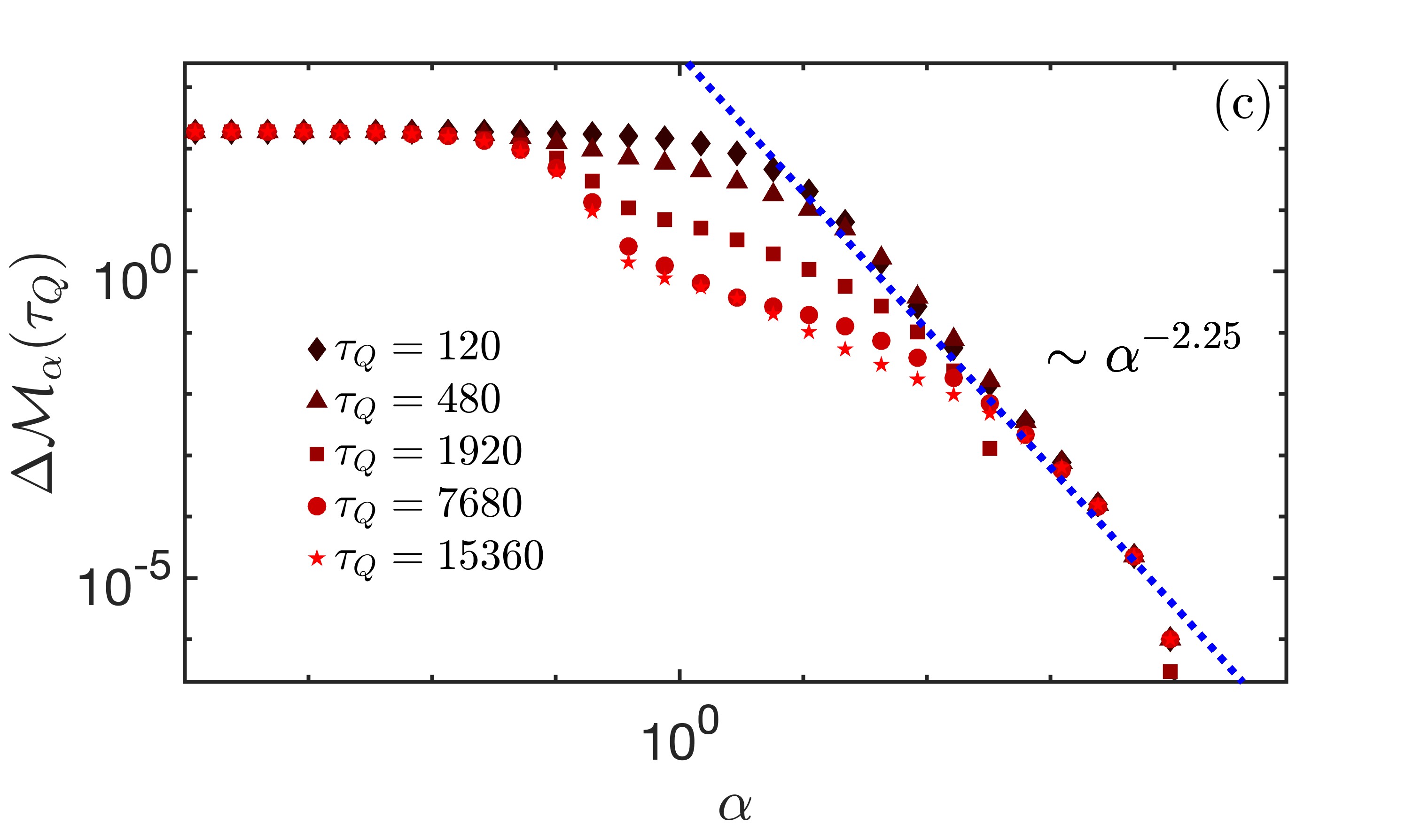}\hfill
  \includegraphics[width=0.47\textwidth,trim={13cm 0 6.5cm 4.5cm},clip]{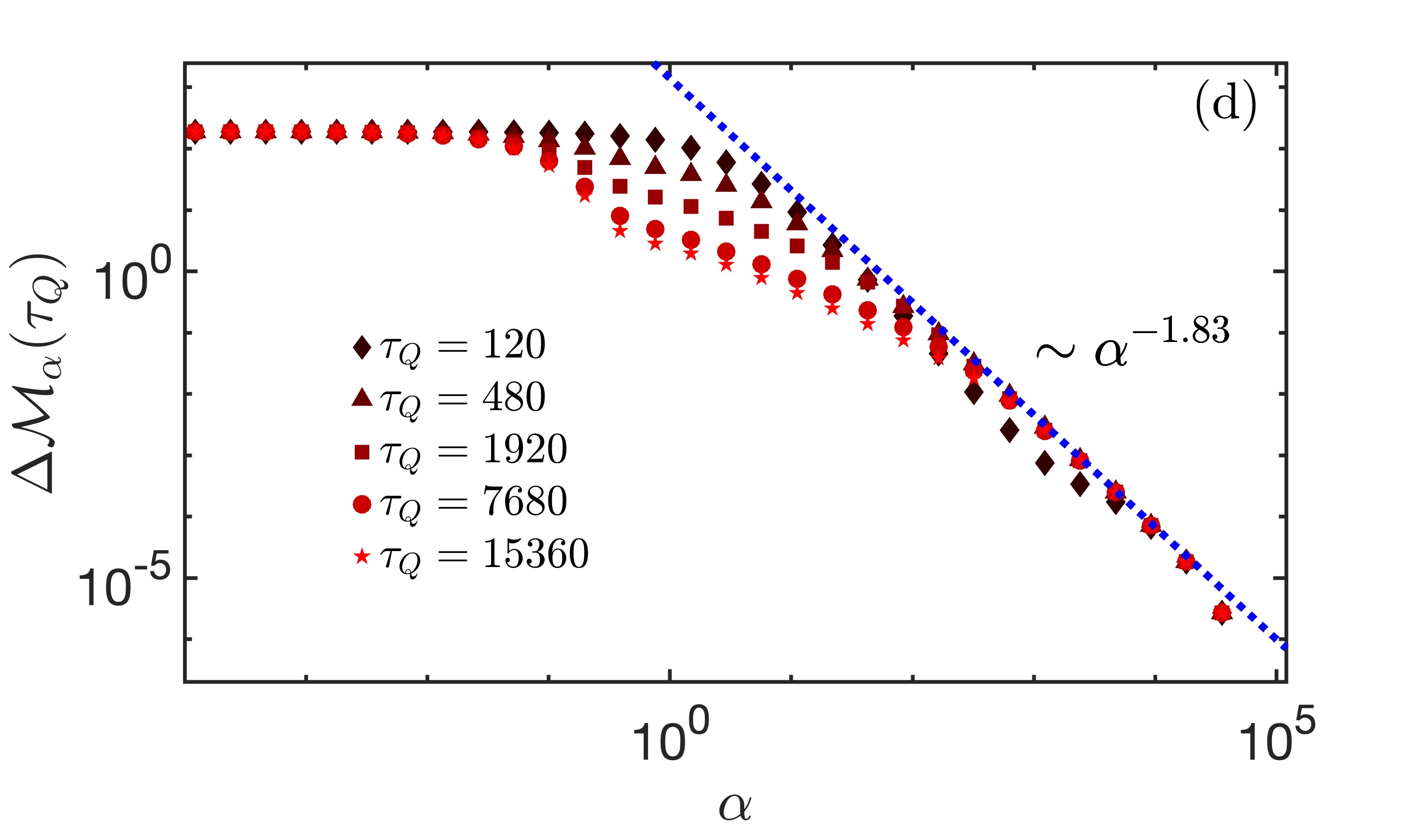}
  \caption{Stabilizer R\'enyi entropies as a function of $\alpha$ for the LRKMs in all the short- and long-range hopping and pairing regimes. (a): $\gamma=5,\,\beta=5$ and (b): $\gamma=1.2,\,\beta=5$ following the same short-range behavior, $\alpha^{-3/2}$ for large $\alpha$. (c): Dynamical scaling regime with $\gamma=1.4,\,\beta=1.6$ and (d): long-range pairing, short-range hopping regime with $\gamma=2.5,\,\beta=1.4$.
  }
  \label{fig:LRKM_Malphaalpha}
\end{figure}

In this section, we provide additional results regarding the stabilizer R\'enyi entropies. From the general expression of Eq.~\eqref{eq: Delta_M_alpha} in the slow driving limit, we consider the large and small $\alpha$ behaviors. Starting with the former
\begin{eqnarray}\label{eq: f_k}
    f(k,\alpha,\tau_Q)&=&\log_2\left[1+\frac{4^\alpha p^\alpha_k(1-p_k)^\alpha\left[(\sin(2\tau_Q))^{2\alpha}+(\cos(2\tau_Q))^{2\alpha}\right]+(\cos\Theta_k-2p_k)^{2\alpha}-\left(\cos\Theta_k\right)^{2\alpha}}{2}\right]\\
    &\approx&\frac{4^\alpha p^\alpha_k(1-p_k)^\alpha\left[(\sin(2\tau_Q))^{2\alpha}+(\cos(2\tau_Q))^{2\alpha}\right]+(\cos\Theta_k-2p_k)^{2\alpha}-\left(\cos\Theta_k\right)^{2\alpha}}{2},
\end{eqnarray}
since in the KZ regime the $p^\alpha_k(1-p_k)^\alpha$ contributions are exponentially suppressed as $\alpha\rightarrow\infty$ the large $\alpha$ asymptotics is governed by the small $k$ expansion of $|1-2p_k|^{2\alpha}$.
Switching to the continuum limit now, one obtains
\begin{eqnarray}
    \Delta\mathcal M_\alpha&\approx&-\frac{L}{4\pi}\tau^{-\frac{1}{2(\beta-1)}}_Q\alpha^{-1} \int_0^\infty\mathrm dx \left[4^\alpha p^\alpha(x)(1-p(x))^\alpha\left[(\sin(2\tau_Q))^{2\alpha}+(\cos(2\tau_Q))^{2\alpha}\right]+(1-2p(x))^{2\alpha}-1\right]\\
    &\approx&-\frac{L}{4\pi}\tau^{-\frac{1}{2(\beta-1)}}_Q\alpha^{-1} \int_0^\infty\mathrm dx (1-2p(x))^{2\alpha}-1\propto -\frac{L}{4\pi}\tau^{-\frac{1}{2(\beta-1)}}_Q\alpha^{-3/2} \int_0^\infty\mathrm dx \left(1-2e^{-C_\beta x^{2(\beta-1)}}\right)^2-1\nonumber\\
    &\approx&\frac{2}{\beta-1}\,\Gamma\!\left(\frac{1}{2(\beta-1)}\right)\,C_\beta^{-\,\frac{1}{2(\beta-1)}}\left(2^{-\,\frac{1}{2(\beta-1)}}-1\right)\frac{L}{4\pi}\tau^{-\frac{1}{2(\beta-1)}}_Q\alpha^{-1-\frac{1}{2(\beta-1)}}\propto\alpha^{-1-1/(2(\beta-1))},\nonumber
\end{eqnarray}
where the first term $(\cos(2\tau_Q))^{2\alpha}
+(\sin(2\tau_Q))^{2\alpha}$ could be neglected as exponentially small compared to the second term.
For small $\alpha\rightarrow 0+$, we use the leading order expansion of Eq.~\eqref{eq: f_k},
\begin{eqnarray}
    & & f(k,\alpha,\tau_Q)=\log_2\left[1+\frac{4^\alpha p^\alpha_k(1-p_k)^\alpha\left[(\sin(2\tau_Q))^{2\alpha}+(\cos(2\tau_Q))^{2\alpha}\right]+(\cos\Theta_k-2p_k)^{2\alpha}-\left(\cos\Theta_k\right)^{2\alpha}}{2}\right]\\
    & & \approx\log_2\frac{3}{2}+\frac{2\alpha}{3\log 2}\left[\log4+\log p_k+\log(1-p_k)+\log(\sin(2\tau_Q))+\log(\cos(2\tau_Q))+\log(\cos\Theta_k-2p_k)-\log(\cos\Theta_k)\right].\nonumber
\end{eqnarray}
As a result, the limiting values of the SRE for small $\alpha$ become
\begin{eqnarray}
    \Delta\mathcal M_\alpha\approx\frac{L}{4}\left[\log_2\frac{3}{2}+O(\tau_Q\alpha)\right],
\end{eqnarray}
where the small $k$ limit for $\cos\Theta_k\approx 1$ was adopted, and the remaining $\tau_Q$  dependent parts have not entered the result as appearing only in bounded oscillatory functions. These features are shown in Fig.~\ref{fig:TFIM_Malpha_alpha}, both for the small and large $\alpha$ limits for the TFIM and in Fig.~\ref{fig:LRKM_Malphaalpha} for the LRKMs.

\clearpage

\end{document}